\documentclass[12pt]{article}
\usepackage[usenames]{color}
\usepackage[dvipdfmx]{graphicx}
\usepackage{amsmath}
\usepackage{amssymb}
\usepackage{amsthm}

\usepackage[dvipdfmx]{hyperref}
\hypersetup{
setpagesize=false,
 bookmarksnumbered=true,%
 bookmarksopen=true,%
 colorlinks=true,%
 linkcolor=blue,
 citecolor=blue,
}
\usepackage{bm}
\usepackage{cite}
\usepackage{framed,color}
\definecolor{shadecolor}{gray}{0.95}
\usepackage{fancybox}
\usepackage{ulem}
\definecolor{shadecolor}{gray}{0.95}
\usepackage{afterpage}
\usepackage{multirow}
\usepackage{slashed}
\numberwithin{equation}{section}

\newcommand{\vev}[1]{\left\langle #1 \right\rangle}
\newcommand{\der}{\partial}
\newcommand{\Tr}{\mbox{\rm Tr}}

\newcommand{\ad}[1]{\mathbf{ad}({#1})}

\newcommand{\Jthe}[4]{\vartheta \begin{bmatrix} {#1}\\{#2} \end{bmatrix} \left({#3}|{#4}\right)}

\begin{document}
\begin{flushright}
\end{flushright}
\begin{center}
{\LARGE\bf Mass spectrum in a six-dimensional $SU(n)$ gauge theory on a magnetized torus}
\vskip 1.4cm
{\large  
Kentaro Kojima$^{a,}$\footnote{E-mail:~kojima@artsci.kyushu-u.ac.jp},
Yuri Okubo$^{b,}$\footnote{E-mail:~okubo.yuri@phys.kyushu-u.ac.jp},
and 
Carolina Sayuri Takeda$^{b,}$\footnote{E-mail:~takeda.carolinasayuri.555@s.kyushu-u.ac.jp}
}\\ \vskip .5cm
{\it
$^a$ Faculty of Arts and Science, Kyushu University, Fukuoka 819-0395, Japan\\%
$^b$ Graduate School of Science, Kyushu University, Fukuoka 819-0395, Japan
}\\
\vskip 1.5cm
\begin{abstract}
We examine six-dimensional $SU(n)$ gauge theories compactified on a two-dimensional torus with a constant magnetic flux background to obtain a comprehensive low-energy mass spectrum. We introduce general background configurations including the magnetic flux and continuous Wilson line phases, consistent with classical equations of motion. Under the standard gauge fixing procedure, the complete mass spectrum in low-energy effective theory for the $SU(n)$ case is newly presented without imposing restrictions on the gauge fixing parameter. Our analysis confirms the inevitable existence of tachyonic modes, which neither depend on the background configurations of Wilson line phases nor are affected by the gauge fixing parameter. Masses for some low-energy modes exhibit dependence on the gauge fixing parameter, and these modes are identified as would-be Goldstone bosons that are absorbed by massive four-dimensional vector fields. We discuss the phenomenological implications associated with stabilization or condensation of the tachyonic states. Various mass spectra and symmetry-breaking patterns are expected with flux backgrounds in the $SU(n)$ case. They are helpful for constructing phenomenologically viable models beyond the standard model, such as gauge-Higgs unification and grand unified theories. 
\end{abstract}
\end{center}

\newpage

\tableofcontents
\bigskip \bigskip

%
\section{Introduction}
\label{sec:intro}
%
The idea of a world with extra spatial dimensions compactified into a
small volume has been a topic of extensive research in elementary
particle physics. In particular, higher-dimensional gauge theories
have been widely studied for seeking physics beyond the Standard Model
(SM). A remarkable feature of higher-dimensional gauge theories is
that extra-dimensional components of gauge fields are four-dimensional
(4D) Lorentz scalars. Hence, without contradicting the 4D Lorentz
invariance, we can consider non-trivial background configurations of
these components, which open up new scenarios for physics beyond the
SM.

A simple possibility is a constant background configuration, namely a
vacuum expectation value (VEV) of extra-dimensional gauge fields. The
VEV of the Higgs scalar field in the SM is a key ingredient and is
forced to have a suitable potential to induce the electroweak symmetry
breaking (EWSB). In addition, in grand unified theories
(GUTs)~\cite{GG}, large gauge symmetries are often supposed to be
broken into the SM gauge symmetry by VEVs of Higgs scalars. Concerning
these issues, the Gauge-Higgs Unification (GHU) models, where Higgs
scalars are identified to light 4D excitations appearing from the
extra-dimensional gauge
field~\cite{manton,fair,hosotani1,Hosotani:1988bm}, have been widely
studied in the contexts of both the EWSB and unified gauge symmetry
breaking~\cite{ Hatanaka:1998yp,Hall:2001zb,Antoniadis:2001cv,Kubo:2001zc,Csaki:2002ur,Burdman:2002se,Gogoladze:2003bb,Scrucca:2003ra,Haba:2004qf,Haba:2004jd,Hosotani:2004ka,Hosotani:2004wv,Haba:2004bh,Lim:2007jv,Kojima:2011ad,Yamashita:2011an,Hosotani:2015hoa,Yamatsu:2015oit,Furui:2016owe,Kojima:2016fvv,Kojima:2017qbt,Hosotani:2017edv,Maru:2019lit,Englert:2019xhz,Angelescu:2021nbp,Nakano:2022lyt,Kojima:2023mew
}. In most of these GHU models, VEVs of extra-dimensional gauge fields
are related to physical degrees of freedom of Wilson line phases
defined with non-contractible cycles on compact
spaces~\cite{hosotani1}. These phases are continuous moduli that
parametrize physical vacua along flat directions of tree-level
potentials for gauge fields. Thus, in many GHU models, Higgs scalars
have flat tree-level potentials and obtain finite effective potentials
through quantum corrections, which is thought to be a result of the
inherent non-locality of the Wilson line phases, as examined in detail
in~\cite{Maru:2006wa,Hosotani:2007kn}. Therefore, GHU models are
considered to have the advantage of clarifying the origin of the
breaking of the electroweak symmetry or unified symmetries.

Another interesting possibility of the background configuration is a
constant magnetic flux provided by extra-dimensional gauge fields. For
the case with two or more extra dimensions, we can turn on the flux
background consistent with classical equations of motion (EOM). The
flux backgrounds may play crucial roles in the compactifications in
string theory associated with moduli stabilization and breaking of
supersymmetry and gauge
symmetry~\cite{Review,Review2,Generations,Bachas}.  In addition, flux
background with toroidal or orbifold compactifications yields chiral
fermions having a generation
structure~\cite{Generations,Bachas,Abe:2008sx,Kobayashi:2010an,Abe:2015yva,Abe:2015mua,Sakamoto:2020pev,Sakamoto:2020vdy,Kobayashi:2022tti,Imai:2022bke},
which is one of the fundamental properties of the fermions in the SM,
in an effective low-energy theory. The flavor structure of quarks and
leptons, such as masses and mixing angles, has been widely examined in
this
setup~\cite{Cremades,Abe:2014vza,Fujimoto:2016zjs,Abe:2016eyh,Kobayashi:2016qag,Buchmuller:2017vut,Buchmuller:2017vho,Neutrino}.

Recently, the mass spectrum of low-energy excitations around flux
backgrounds has gained much attention and has been examined in detail,
including quantum
corrections~\cite{Buchmuller,Ghilencea:2017jmh,B1,xi,Honda:2019ema,Maru,Maru2,Akamatsu}. In
six-dimensional (6D) models with a $T^2$ compactification, massless
scalar excitations appear from the extra-dimensional gauge field that
provides the magnetic flux background. The massless scalars are
identified as the Nambu-Goldstone (NG) bosons associated with the
translational symmetry broken by the flux background. Recent studies
explicitly showed that exact cancellation occurs in one-loop
contributions to masses of these
scalars~\cite{Buchmuller,Ghilencea:2017jmh,B1,xi}. Toward applications
to phenomenologically viable GHU models, finite masses for the
pseudo-NG modes are investigated with the help of explicit breaking of
the translational symmetry through interaction terms on
$T^2$~\cite{Maru2} or fixed points on
$T^2/\mathbb Z_2$~\cite{Maru:2023esr}.

A non-vanishing flux background taking a direction in the space of a
simply-connected gauge group makes some of the 4D parts of gauge
fields massive and must induce a spontaneous gauge symmetry breaking
at low energy. In addition, such backgrounds have been discussed to
accompany tachyonic excitations at a low-energy
regime~\cite{Bachas,Cremades,Buchmuller}, related to the
Nielsen-Olesen instabilities~\cite{Nielsen:1978rm}. Condensation of
these tachyonic modes seems to have an impact on vacuum structure, as
examined in superstring
models~\cite{Buchmuller:2019zhz,Buchmuller:2020nnl}, and may give rich
phenomena such as further symmetry breaking in a more general field
theoretical setup. Despite these phenomenologically interesting
features, flux backgrounds associated with an $SU(n)$ gauge group have
been less studied, and more comprehensive studies are needed to
explore phenomenologically viable models based on the flux background.

This work mainly focuses on clarifying complete tree-level mass
spectra appearing in low-energy effective theories with flux
backgrounds in a general $SU(n)$ case. We study 6D non-supersymmetric
gauge theory compactified on $T^2$ in detail. We examine the classical
EOM to obtain consistent background configurations including both the
magnetic flux and Wilson line phases.  With the consistent background
configurations, boundary conditions for fields associated with the
discrete translations on $T^2$ are studied. The background
configurations and the boundary conditions simultaneously change under
gauge transformations. With the help of gauge transformations, the
Wilson line phases are removed from the background, and the mass
spectrum of low-energy 4D excitations is discussed. We explicitly
perform a standard $R_\xi$ gauge fixing, keeping the calculations as
general as possible, and discuss physically relevant excitations at a
low-energy regime.  Under a mode expansion of six-dimensional fields,
non-trivial mixing among four-dimensional modes appears in their mass
terms depending on the gauge fixing terms. We clarify the expressions
for the masses depending on the gauge parameter. We confirm that
tachyonic scalars inevitably appear in this setup. In addition, some
of the scalar excitations are identified as would-be Goldstone modes
that are absorbed by 4D vector fields, which become massive due to the
gauge symmetry breaking triggered by the flux background. We also
discuss the phenomenological implications associated with
stabilization or condensation of the tachyonic states. Various mass
spectra and symmetry-breaking patterns are expected with the flux
background for a simply-connected gauge group and are interesting for
constructing phenomenologically viable models beyond the standard
model, such as GHU and GUT.

The structure of this paper is as follows. In section~\ref{sec:sunym},
we present our definitions and the basic concepts of $SU(n)$ gauge
theories. The boundary conditions and consistency conditions for the
gauge field are also defined. Taking these conditions into
consideration, we obtain a background configuration solution in
section~\ref{sec:bcT2}. In section~\ref{sec:para}, we discuss a
convenient parametrization of this background. In
section~\ref{sec:treeLag}, we examine the Yang-Mills Lagrangian in
this setup, fixing the gauge and obtaining the explicit expressions
for the quadratic terms. From these terms, we could compute the
tree-level mass spectrum for all of the fields, as done in
section~\ref{sec:treemass}. Finally, phenomenological implications are
discussed in section~\ref{sec:pheno}, followed by our
conclusions. Calculation details can be found in the appendices.

%
\section{$SU(n)$ gauge theories on ${\cal M}^4\times T^2$}
\label{sec:sunym}
%
We study gauge theories on ${\cal M}^4\times T^2$, where ${\cal M}^4$
and $T^2$ are the Minkowski spacetime and a two-dimensional torus,
respectively. We denote coordinates by $x^\mu$ $(\mu=0,1,2,3)$ on
${\cal M}^4$ and by $x^5$ and $x^6$ on $T^2$. We also use $x^M$
($M=0,1,2,3,5,6$) and $x^m$ $(m=5,6)$. A two-dimensional torus $T^2$
is given by imposing the identification for $x^5$ and $x^6$ as
\begin{align}
  (x^5,x^6)\sim \left(x^5+L(n_5+n_6\tau_{\rm R}),x^6+Ln_6\tau_{\rm I}\right),
  \qquad n_5,n_6\in\mathbb Z, \qquad \tau_{\rm I}>0,
 \label{defT2tor1}
\end{align}
where $\tau=\tau_{\rm R}+i\tau_{\rm I}$
$(\tau_{\rm R},\tau_{\rm I}\in\mathbb R)$ is a complex parameter
describing the moduli space of the two-dimensional torus. The size of
$T^2$ is parametrized by $L$, where the volume of $T^2$ is given by
${\cal V}_{T^2}=L^2 \tau_{\rm I}$. In the following, we take $L=1$
without loss of generality.

It is convenient to use complex coordinates defined as
\begin{align}\label{zdef1}
  z=x^5+ix^6,
\qquad \bar z=x^5-ix^6, 
  \qquad \der_z={1\over 2}(\der_5-i\der_6), 
  \qquad \bar \der_z={1\over 2}(\der_5+i\der_6), 
\end{align}
where $ \der_M={\der /\der x^M}$.  Then, eq.~\eqref{defT2tor1} is
expressed by $z \sim z + n_5 + n_6\tau$. Let us define the translation
operators ${\cal T}_p$ $(p=5,6)$ as
\begin{align}
  {\cal T}_5z=z+1, \qquad {\cal T}_6 z = z +\tau. 
\end{align}
Note that the direction of the translation generated by ${\cal T}_6$
is different from the one of $x^6$ for $\tau_{\rm R}\neq 0$, whereas
the directions of ${\cal T}_5$ and $x^5$ coincide with each
other. Using these operators, eq.~\eqref{defT2tor1} is rewritten by
\begin{align}\label{zid2}
  z\sim {\cal T}_6^{n_6}{\cal T}_5^{n_5}z.
\end{align}

We consider gauge theories on ${\cal M}^4\times T^2$ with the gauge
group $SU(n)$, whose Lie algebra is denoted by $su(n)$. The action is
given by the 6D volume integral of a Lagrangian density ${\cal L}_6$
of an $SU(n)$ gauge theory. We demand that ${\cal L}_6$ is invariant
under 6D Lorentz transformations and $SU(n)$ gauge transformations.

Let $\bm A_M\in su(n)$ be a gauge field, which is a function of
$x^M$. In order to expand the gauge field as $\bm A_M=A_M^at_a$
$(A_M^a\in \mathbb R,\ a=1,\dots,n^2-1)$, we introduce generators
$t_a\in su(n)$, which span the vector space $su(n)$.  Hereafter, we
imply summations over the same upper and lower indices.  We refer to
$A_M^a$ as a component field. We also introduce the covariant
derivative $\bm D_M$ as
\begin{align}
  \bm D_M=\der_M-ig \bm A_M,
\end{align}
where $g$ is a gauge coupling constant. 

We first discuss the pure Yang-Mills theory; matter fields are
discussed in section~\ref{sec:pheno}. The Lagrangian is given by
\begin{align}\label{Lpym}
  {\cal L}_{\rm YM}=-{1\over 2} \eta^{MM'}\eta^{NN'}\Tr (\bm F_{MN}\bm F_{M'N'}),
\end{align}
where $\bm F_{MN}$ is the field strength, and $\eta^{MN}$ is the
metric of the 6D spacetime.  The trace is implied to be taken in a
representation space of $su(n)$.  The field strength is given by
\begin{align}\label{defFmn}
  \bm F_{MN}={i\over g}[\bm D_M,\bm D_N]=\der_M\bm A_N-\der_N\bm A_M -ig[\bm A_M,\bm A_N],
\end{align}
where $[A,B]=AB-BA$, and the metric is defined to be
$(\eta^{MN})={\rm diag}(-1,1,\dots,1)$.  Namely, $x^M$ are orthogonal
coordinates.  We also use $(\eta^{\mu\nu})={\rm diag}(-1,1,1,1)$ for
the Minkowski part. Note that $\bm F_{MN}$ is written by
$F_{MN}^at_a\in su(n)$.

The Lagrangian in eq.~\eqref{Lpym} is invariant under a gauge
transformation, which is defined by
\begin{align}\label{gtam1}
  \bm A_M\to \bm A_M^{\rm gt}=\bm \Lambda \bm A_M \bm \Lambda^{-1}+{i\over g}\bm \Lambda\der_M
  \bm \Lambda^{-1}, 
\end{align}
where $\bm \Lambda\in SU(n)$ is a function of $x^M$ and is called a
gauge transformation function.  From eq.~\eqref{gtam1}, one finds
$\bm D_M\to \bm D_M^{\rm gt}=\bm \Lambda \bm D_M \bm \Lambda^{-1}$ and
$\bm F_{MN}\to \bm F_{MN}^{\rm gt}=\bm \Lambda \bm F_{MN}
\bm\Lambda^{-1}$, under which the gauge invariance of the Lagrangian
in eq.~\eqref{Lpym} is manifested.

For later convenience, we define
\begin{align} \label{defaz1}
  \bm A_z&={1\over 2}(\bm A_{5}-i\bm A_{6}), &
  \bar {\bm A}_z
  &={1\over 2}(\bm A_{5}+i\bm A_{6}), \\
 \bm D_z&={1\over 2}(\bm D_5-i\bm D_6)=\der_z-ig \bm A_z, &
 \bar {\bm D}_z&={1\over 2}{(\bm D_5+i\bm D_6)}
 =\bar \der_z-ig \bar {\bm A}_z.
\end{align}
From the above and eq.~\eqref{zdef1}, one naturally defines
\begin{align}
  \bm F_{z\bar z}=\der_z\bar {\bm A}_z-\bar \der_{z}\bm A_z-ig[\bm A_z,\bar {\bm A}_{z}]
  ={i\over 2}\bm F_{56},
  \qquad \bm F_{\bar z z}=-\bm F_{z\bar z}, \qquad
  \bm F_{zz}=\bm F_{\bar z \bar z}=0.
\end{align}
Similar notations are used for component fields, e.g.,
$F_{z\bar z}^a=iF_{56}^a/2$.

Since we consider compact extra dimensions, boundary conditions for
fields have to be specified to define gauge theories. In view of the
identification in eq.~\eqref{zid2}, we require that
$\bm A_M(x^\mu,{\cal T}_6^{n_6}{\cal T}_5^{n_5}z)$ is equal to
$\bm A_M(x^\mu,z)$ up to a gauge transformation as a  sufficient
condition to make the pure Yang-Mills Lagrangian single-valued on
$T^2$. Hence, the boundary conditions are defined as
\begin{align}\label{bcdef1}
  \bm A_M({\cal T}_pz)&=T_p(z)\bm A_M(z)T_p^\dag(z)+{i\over g} T_p(z)\der_MT_p^\dag(z),
\end{align}
where we have introduced $T_p(z)\in SU(n)$ $(p=5,6)$.  As a shorthand
notation, $x^\mu$ was suppressed, and $\bm A_M(x^\mu,z)$ and
$T_p(x^\mu,z)$ are written as $\bm A_M(z)$ and $T_p(z)$, respectively.
A similar notation is used for others that depend on $x^M$. The
boundary conditions are specified by $T_p$, which we refer to as twist
matrices hereafter. The twist matrices generally depend on $x^M$.

Different twist matrices can be physically equivalent since the twist
matrices $T_p$ depend on the choice of
gauge~\cite{Hosotani:1988bm,Haba:2002py,Haba:2003ux,Kawamura:2022ecd}. From
eq.~\eqref{gtam1}, one sees that
\begin{align}\label{deftidash}
  \bm A_M^{\rm gt}({\cal T}_pz)&=T_p^{\rm gt}(z)\bm A_M^{\rm gt}(z)T_p^{{\rm gt}\dag}(z)+{i\over g} T_p^{\rm gt}(z)\der_MT_p^{{\rm gt}\dag}(z),
\end{align}
where 
\begin{align}\label{gttwistdef}
  T_p^{\rm gt}(z)=  \bm \Lambda({\cal T}_p(z))T_p(z)\bm \Lambda^\dag(z). 
\end{align}

The gauge field at $z+1+\tau$ is written by
\begin{align}
  \bm A_M(z+1+\tau)=\bm A_M(\mathcal{T}_6(z+1))=
  \bm A_M(\mathcal{T}_5(z+\tau)).
\end{align}
From the above, we obtain 
\begin{align}\label{con_kg}
  [T_\square(z),\bm A_M(z)]={i\over g}\der_M T_\square(z),
\qquad 
  T_\square(z)
  =T_6^\dag(z)T_5^\dag(z+\tau)T_6(z+1)T_5(z), 
\end{align}
which is regarded as a consistency condition for the twist matrices.

As seen below, there appear to be additional consistency conditions
for the twist matrices if non-trivial background configurations for
the gauge field exist.  In the next section, we discuss explicit forms
of the twist matrices taking background configurations for the gauge
field into account.

\section{Consistency between background configurations and boundary conditions}
\label{sec:bcT2}
Let us introduce non-trivial background configurations for the gauge
field $\bm A_M$. For consistency, we demand that the backgrounds
satisfy the classical EOM. From the Lagrangian in eq.~\eqref{Lpym}, we
find that the EOM is given by
\begin{align}
  \bm D^M\bm F_{MN}=  \left(\der^M-ig\ad{\bm A^M}\right)\bm F_{MN}=0,
\end{align}
where we have introduced the notation $\ad{X}Y=[X,Y]$.  In the
following, we replace
\begin{align}\label{repl1}
  \bm A_M(z)\to \bm B_M(z) + {\bm A}_M(z), 
\end{align}
in the Lagrangian in eq.~\eqref{Lpym}. On the right-hand side,
$\bm B_M$ and ${\bm A}_M$ are referred to as the background and the
fluctuation around the background configuration, respectively.
Imposing $\bm B_\mu=0$ and $\der_\mu \bm B_M=0$, we can explicitly
keep the 4D Lorentz invariance. Then, non-trivial background
configurations can be given by $\bm B_m\neq 0$, which generally depend
on the torus coordinates.

It is convenient to introduce the background covariant derivative
${\cal D}_m$ and the field strength ${\cal F}_{mn}$ as
\begin{align}\label{bgdfdef1}
  {\cal D}_m=\der_m-ig\ad{\bm B_m},\qquad
    {\cal F}_{mn}=\der_m\bm B_n-\der_n\bm B_m-ig[\bm B_m,\bm B_n].
\end{align}
According to the above discussion, we require that
\begin{align}  \label{eombg1}
{\cal D}^m  {\cal F}_{mn}=0, 
\end{align}
to satisfy the consistency condition for the background field
$\bm B_m$.

A solution to eq.~\eqref{eombg1} is written as 
\begin{gather}
  \label{bgconst1}
  \bm B_5(z)=\bm v_5-\left(1+\gamma\right) \bm fx^6/2, \qquad
  \bm B_6(z)=\bm v_6+\left(1-\gamma\right) \bm fx^5/2, \\
  [\bm v_5,\bm v_6]=[\bm v_m,\bm f]=0,
  \label{bgconst12}
\end{gather}
where $\bm v_m,\bm f\in su(n)$ and $\gamma\in \mathbb R$ are
constants.  With this background configuration, one sees that
${\cal F}_{56}=\bm f$. We call $\bm f$ a constant magnetic flux,
whereas $\bm v_m$ are referred to as continuous Wilson line phases. As
discussed in the next section, allowed values of $\bm f$ are
quantized. On the other hand, continuous variables $\bm v_m$ are
related to the flat directions in the tree-level potential for the
gauge field obtained from eq.~\eqref{Lpym}.  The parameter $\gamma$ is
introduced for clarity of our discussions and is independent of
${\cal F}_{56}$.  In the literature, $\gamma=\pm 1$ and $\gamma=0$ are
often called the Landau and the symmetric gauge, respectively.

The background $\bm B_m$ changes by gauge transformations. Using the
gauge transformation in eq.~\eqref{gtam1} with a constant
$\bm \Lambda$, we can diagonalize $\bm f$ in a representation space of
$su(n)$. Then, the last equality in eq.~\eqref{bgconst12} implies that
we can also diagonalize $\bm v_m$ keeping $\bm f$ diagonal. Thus,
without loss of generality, we can expand $\bm f$ and $\bm v_m$ by
Cartan generators. We discuss the explicit forms of the $su(n)$
generators in the next section.

Considering the background configuration in eq.~\eqref{bgconst1}, we
examine the boundary conditions for the gauge field in
eq.~\eqref{bcdef1}. The background $\bm B_M$ and the twist matrices
$T_p$ must satisfy the relation
\begin{align}\label{bgtwist1}
    \bm B_M({\cal T}_pz)&=T_p(z)\bm B_M(z)T_p^\dag(z)+{i\over g} T_p(z)\der_MT_p^\dag(z), 
\end{align}
whereas the fluctuations satisfy
\begin{align}\label{bcfluc1}
  {\bm A}_M({\cal T}_pz)&=T_p(z){\bm A}_M(z)T_p^\dag(z).
\end{align}
From eq.~\eqref{bgtwist1} for $M=\mu$, one sees that $\der_\mu
T_p=0$. On the other hand, from eq.~\eqref{bgtwist1} for $M=n$, we
obtain
\begin{align}\label{twistcon21}
\der_5T_5&=ig[\bm B_5,T_5],\\
\der_6T_5&=ig[\bm B_6,T_5]+ig\left(1-\gamma\right)\bm f T_5/2,\\
\der_5T_6&=ig[\bm B_5,T_6]-ig\left(1+\gamma\right)\tau_{\rm I} \bm f T_6/2,\\
  \der_6T_6&=ig[\bm B_6,T_6]+ig\left(1-\gamma\right)\tau_{\rm R}\bm f T_6/2,
             \label{twistcon24}
\end{align}
where $\bm B_m$ and $T_p$ are defined at $z$. The twist matrices must
satisfy these conditions in addition to eq.~\eqref{con_kg}.

A solution to eqs.~\eqref{twistcon21}--\eqref{twistcon24} with the
background configuration in eq.~\eqref{bgconst1} is given by
\begin{align}\label{twistLanG}
  T_5(z)= e^{ig(1-\gamma)\bm fx^6/2}\tilde T_5,\qquad 
  T_6(z)= e^{ig\{-(1+\gamma) \tau_{\rm I}\bm fx^5/2+(1-\gamma) \tau_{\rm R}\bm fx^6/2\}}\tilde T_6, \end{align}
where we have introduced constant matrices $\tilde T_p$ that satisfy $[\bm f,\tilde T_p]=[\bm v_n,\tilde T_p]=0.$
Then, $T_\square(z)$ defined in eq.~\eqref{con_kg} becomes
\begin{align}
    T_\square(z)=e^{-ig\mathcal{V}_{T^2}\bm f}\tilde T_6^\dag \tilde T_5^\dag
  \tilde T_6\tilde T_5.
\end{align}
From the right-hand side of the above equation, one sees that
$T_\square(z)$ is a constant matrix. Hence, eq.~\eqref{con_kg} is
reduced to $[T_\square,\bm A_M(z)]=0$.  The general solution of this
constraint is $T_\square\in \mathbb Z_n$, where
$\mathbb Z_n\subset SU(n)$ is the center subgroup of $SU(n)$. Thus, we
obtain
\begin{align}\label{con_kg2}
  e^{-ig\mathcal{V}_{T^2}\bm f}\tilde T_6^\dag \tilde T_5^\dag
  \tilde T_6\tilde T_5=e^{2\pi i \tilde n/n}\bm I, \qquad \tilde n\in \mathbb Z,
\end{align}
where $I\in SU(n)$ is the identity operator. The integer $\tilde n$
modulo $n$ is referred to as the 't~Hooft flux~\cite{tHooft:1979rtg}.

Although there are interesting possibilities of non-trivial choices
for $\tilde T_p$ and $\tilde n$ in eq.~\eqref{con_kg2}, we restrict
our attention to the simplest $\tilde n=0$ and $\tilde T_p=\bm I$ in
the following discussions. Then, from eq.~\eqref{con_kg2}, we obtain
\begin{align}\label{fluxq1}
  e^{ig\mathcal{V}_{T^2}\bm f}=\bm I, 
\end{align}
which gives a quantization condition for the flux $\bm f$. We will
examine this condition in the next section.

Both the background $\bm B_m$ and the twist matrices $T_p$ have gauge
dependence. A notable fact is that the constant terms $\bm v_m$ in
eq.~\eqref{bgconst1} can be eliminated by a gauge transformation with
$\bm \Lambda=e^{-ig(\bm v_5x^5+\bm v_6x^6)}$. From eqs.~\eqref{gtam1}
and~\eqref{gttwistdef}, after the gauge transformation, one finds
\begin{gather}\label{bmnewg1}
  \bm B_5(z)=-(1+\gamma)\bm fx^6/2, \qquad 
  \bm B_6(z)=(1-\gamma)\bm fx^5/2, \\
  T_5(z)=e^{-ig\bm v_5}e^{ig(1-\gamma)\bm fx^6/2}, \qquad
  T_6(z)=e^{-ig(\tau_{\rm R}\bm v_5+\tau_{\rm I}\bm v_6)}
  e^{ig\{-(1+\gamma) \tau_{\rm I}\bm fx^5/2+(1-\gamma) \tau_{\rm R}\bm fx^6/2\}},
  \label{bmnewg2}
\end{gather}
where we have renamed $\bm B_m^{\rm gt}$ and $T_p^{\rm gt}$ as
$\bm B_m$ and $T_p$ to simplify the notation.  From the above, it is
clear that non-trivial values of the continuous Wilson line phases
$\bm v_m$ can be treated as some part of the twist
matrices~\cite{Hall:2001tn}. We choose this gauge in the following
discussions.

\section{Parametrization of background gauge fields}
\label{sec:para}
Considering the background configuration in eq.~\eqref{bmnewg1}, let
us discuss the parametrization of $\bm B_m$ by $su(n)$ generators. Let
$\{t_a\}$ $(a=1,\dots,n^2-1)$ be a set of the $su(n)$ generators. It
is convenient to use the Cartan-Weyl basis, writing
$\{t_a\}=\{H_k\}\cup \{E_{\bm \alpha}\}$, where $\{H_k\}$
$(k=1,\dots,n-1)$ are Cartan generators, and $E_{\bm \alpha}$ is a
step operator associated to a root vector $\bm \alpha$. The Cartan
generators are Hermitian, $H_k^\dag=H_k$, and step operators
$E_{\bm \alpha}$ satisfy $E_{\bm \alpha}^\dag=E_{-\bm \alpha}$. They
obey the commutation relations
\begin{align}\label{CWcom1}
  [H_k,H_\ell]=0, \qquad
  [H_k,E_{\bm \alpha}]=\alpha_{[k]}E_{\bm \alpha}, 
\end{align}
where $\alpha_{[k]}\in \mathbb R$ is the $k$--th component of the root
vector ${\bm \alpha}$.  As noted before, $\bm f$ and $\bm v_m$ are
expanded by $su(n)$ Cartan generators:
\begin{align}
  \bm f=f^kH_k, \qquad \bm v_m=v_m^kH_k, \qquad f^k,v_m^k\in \mathbb R.
\end{align}

To be more concrete, we fix a basis of the generators in a
representation space of $su(n)$. Let us denote the fundamental
representation of $H_k$ by $\hat H_k$ and take
\begin{gather}\label{Hnormfund1}
\hat H_1={\rm diag}(1,-1,0,\dots,0), \quad 
  \hat H_2={\rm diag}(0,1,-1,0,\dots,0), \quad
  \dots, \\ 
\hat H_{n-2}={\rm diag}(0,\dots,0,1,-1,0), \quad 
\hat H_{n-1}={\rm diag}(0,\dots,0,1,-1).
\label{Hnormfund2}
\end{gather}
In the following discussions, we identify any operators of $su(n)$ as
their fundamental representation matrices for simplicity.

Let us examine the quantization condition of the constant magnetic
flux in eq.~\eqref{fluxq1}. It is convenient to introduce the unit
strength of the magnetic flux $\hat f={2\pi/(g{\cal V}_{T^2})}$. Then,
using the explicit forms of $\hat H_k$ in eqs.~\eqref{Hnormfund1}
and~\eqref{Hnormfund2}, we obtain a general solution to
eq.~\eqref{fluxq1} as
\begin{align}
  f^k={2\pi\over g{\cal V}_{T^2}} N^k
  =\hat fN^k, \qquad N^k\in \mathbb Z.
  \label{qcondhatb1}
\end{align}
The flux $\bm f$ in the fundamental representation is expressed by
\begin{align}
  \label{ffund1}
  \bm f=f^k\hat H_k=\hat f~{\rm diag}(N^1,N^2-N^1,\dots,N^{n-1}-N^{n-2},-N^{n-1}).
\end{align}

For later convenience, we discuss the step operators in the
fundamental representation. We define $n^2-1-(n-1)=n(n-1)$ step
operators. To express them, it is convenient to introduce the basis
matrices ${\hat e}_{ij}$, whose $(i',j')$ element
$(\hat e_{ij})_{i'j'}$ is given by
$(\hat e_{ij})_{i'j'}=\delta_{ii'}\delta_{jj'}$, where $\delta_{ii'}$
is the Kronecker delta.  We can denote the $n(n-1)$ step operators in
the fundamental representation by
\begin{align}\label{Eijdef1}
  E_{ij}^{(+)}={\hat e}_{ij}, \qquad 
  E_{ij}^{(-)}={\hat e}_{ji}, \qquad 1\leq i<j\leq n.
\end{align}
The Cartan generators are written by
$\hat H_k=\hat e_{kk}-\hat e_{k+1\,k+1}$ with these basis
matrices. From the definition of the fundamental representation of the
generators, one sees that
\begin{gather}\label{trgen1}
  \Tr[\hat H_k\hat H_\ell]=(M^{su(n)})_{k\ell},
  \qquad \Tr[E_{ij}^{(+)}E_{i'j'}^{(-)}]=\delta_{ii'}\delta_{jj'}, \\
\Tr[\hat H_{k}E_{ij}^{(\pm)}]=
  \Tr[E_{ij}^{(+)}E_{i'j'}^{(+)}]
  =\Tr[E_{ij}^{(-)}E_{i'j'}^{(-)}]=0,
  \label{trgen2}
\end{gather}
where $(M^{su(n)})_{k\ell}$ is the $(k,\ell)$ element of the $su(n)$
Cartan matrix
\begin{align}
M^{su(n)}=
  \begin{pmatrix}
      2&-1&0&\cdots&&&0\\
      -1&2&-1&0&\cdots&&0\\
      &&&\ddots
      &&&\\
      0&\cdots&&0&-1&2&-1\\
      0&\cdots&&&0&-1&2
  \end{pmatrix}.
\end{align}

In the following discussions, commutation relations between the
generators often appear, such as
\begin{align}\label{comHkEij1}
  [\hat H_k,E_{ij}^{(\pm)}]=\pm (\delta_{ki}-\delta_{k+1\,i}-\delta_{kj}+\delta_{k+1\,j})E_{ij}^{(\pm)}, 
\end{align}
where we note that, on the right-hand side, the indices are not
summed. The eigenvalue
$\pm (\delta_{ki}-\delta_{k+1\,i}-\delta_{kj}+\delta_{k+1\,j})$ is the
$k$--th component of the root vector corresponding to
$E_{ij}^{(\pm)}$. From eqs.~\eqref{ffund1} and~\eqref{comHkEij1}, we
obtain
\begin{align}\label{fecom1}
  [\bm f,E_{ij}^{(+)}]
  &= \hat f (N^i-N^{i-1}-N^{j}+N^{j-1})E_{ij}^{(+)}
    \equiv \hat f \tilde N^{ij}E_{ij}^{(+)}, \\\label{vecom1}
  [\bm v_m,E_{ij}^{(+)}]
  &=
     (v_m^i-v_m^{i-1}-v_m^{j}+v_m^{j-1})E_{ij}^{(+)}
    \equiv \tilde v_{m}^{ij}E_{ij}^{(+)}, 
\end{align}
where we have defined
\begin{align}\label{deftilN}
  \tilde N^{ij} =N^i-N^{i-1}-N^{j}+N^{j-1}, \qquad
  \tilde v_{m}^{ij}=v_m^i-v_m^{i-1}-v_m^{j}+v_m^{j-1}.
\end{align}
We note that we have implied $N^{i-1},v_m^{i-1}=0$ for $i=1$ and
$N^{i},v_m^{i}=0$ for $i=n$.  We also have
$[\bm f,E_{ij}^{(-)}]=-\hat f \tilde N^{ij}E_{ij}^{(-)}$,
$[\bm v_m,E_{ij}^{(-)}]=-\tilde v_{m}^{ij}E_{ij}^{(-)}$, and
$[\bm f,\hat H_k]=[\bm v_m,\hat H_k]=0$.

\section{Lagrangian around flux backgrounds}
\label{sec:treeLag}

\subsection{Lagrangian for the fluctuations}
\label{sec:fluc}
Let us examine the Lagrangian in eq.~\eqref{Lpym} with the background
configurations and boundary conditions in eqs.~\eqref{bmnewg1}
and~\eqref{bmnewg2}. We aim to study the tree-level mass spectrum in
the effective 4D theory, which describes physics at a sufficiently
lower energy scale than the compactification scale $1/L$. Since the
quadratic terms of the fields in the Lagrangian are relevant to the
mass spectrum, we mainly focus on them here.

As discussed, we take the replacement in eq.~\eqref{repl1} in the
Lagrangian in eq.~\eqref{Lpym}. We use the background covariant
derivative and the field strength defined in eq.~\eqref{bgdfdef1},
which are generalized to
\begin{align}
  {\cal D}_M=
    \begin{cases}
        {\cal D}_\mu=\der_\mu,\\
        {\cal D}_m=\der_m-ig\ad{\bm B_m},
    \end{cases} \qquad 
      {\cal F}_{MN}=
    \begin{cases}
        {\cal F}_{\mu M}=0,\\
        {\cal F}_{mn}=\epsilon_{mn}\bm f,
    \end{cases}
\end{align}
where $\epsilon_{56}=-\epsilon_{65}=1$.  Then, the field strength in
eq.~\eqref{defFmn} is rewritten as 
\begin{gather}\label{FMNrewrite}
    \bm F_{MN}={\cal F}_{MN}+{\cal D}_M\bm A_N-{\cal D}_N\bm
    A_M-ig[\bm A_M,\bm A_N],
\end{gather}
where the first term on the right-hand side is a constant.

The Lagrangian in eq.~\eqref{Lpym} is decomposed into
\begin{align}
  {\cal L}_{\rm YM}=-{1\over 2}\sum_{p=0}^4\Tr[L^{(p)}], 
\end{align}
where $L^{(p)}$ contains $p$-th polynomials of the fluctuations. We find 
\begin{align}
L^{(0)}&={\cal F}_{MN}{\cal F}^{MN},\\
L^{(1)}&=4{\cal F}^{MN}{\cal D}_M\bm A_N ,\\
L^{(2)}&=2{\cal D}_M\bm A_N({\cal D}^M\bm A^N-{\cal D}^N\bm A^M)
          -2ig{\cal F}^{MN}[\bm A_M,\bm A_N] ,\\ 
L^{(3)}&=-4ig ({\cal D}_M\bm A_N) [\bm A^M,\bm A^N], \\
  L^{(4)}&=-g^2[\bm A^M,\bm A^N][\bm A_M,\bm A_N].
\end{align}
The constant terms $L^{(0)}$ contribute to the cosmological constant,
which is irrelevant to our present analysis.  The linear terms
$L^{(1)}$ vanish in the action.\footnote{The linear terms $L^{(1)}$
  yield two types of contributions in the Lagrangian. One of them is
  the term proportional to $\epsilon^{mn}\der_m\Tr[\bm f \bm A_n]$.
  Since $\Tr[\bm f \bm A_n]$ is periodic under the shift generated by
  ${\cal T}_p$, this surface term vanishes in the action. Another type
  is the term proportional to $\Tr[\bm f[\bm B_m,\bm A_n]]$. Using the
  parametrization of $\bm B_m$ and the relations in
  eqs.~\eqref{CWcom1} and~\eqref{trgen2}, we find that such terms
  vanish due to $\Tr[\hat H_{k}E_{ij}^{(\pm)}]=0$.} Such terms induce
tadpoles and must vanish with background configurations consistent
with the EOM. The quadratic terms $L^{(2)}$ describe the tree-level
mass spectrum in the 4D effective theory.  We study them in detail
just below. From the trilinear and quartic terms $L^{(3)}$ and
$L^{(4)}$, one finds
\begin{align}\notag
 -{1\over 2} \Tr[L^{(3)}]
  &=2ig\Tr[(\der_\mu \bm A_\nu)[\bm A^\mu,\bm A^\nu]]
    +2ig\Tr[(\der_\mu \bm A_m)[\bm A^\mu,\bm A^m]]\\\label{L31}
  & \qquad  +2ig\Tr[({\cal D}_m \bm A_\mu)[\bm A^m,\bm A^\mu]]
    +2ig\Tr[({\cal D}_m \bm A_n)[\bm A^m,\bm A^n]],\\
 -{1\over 2} \Tr[L^{(4)}]
  &= {g^2\over 2}\Tr[([\bm A_\mu,\bm A_\nu])^2]
    +g^2\Tr[([\bm A_\mu,\bm A_m])^2 ] +{g^2\over 2}\Tr[([\bm A_m,\bm A_n])^2],
    \label{L41}
\end{align}
which give the interactions between the fluctuations. 

For the quadratic terms, a straightforward calculation shows
\begin{align}\label{quad10}
  -{1\over 2} \Tr[L^{(2)}]
  &= {\cal L}^{(2)}_1+{\cal L}^{(2)}_2+{\cal L}^{(2)}_3,\\\label{quad11}
  {\cal L}^{(2)}_1
  &=\Tr[-(\der_\mu \bm A_\nu)^2-({\cal D}_m\bm A_\mu)^2+(\der_\mu \bm A_\nu)(\der^\nu \bm A^\mu)],\\\label{quad12}
  {\cal L}^{(2)}_2
  &=  \Tr\left[-(\der_\mu \bm A_m)^2-({\cal D}_m\bm A_n)^2+({\cal D}_m\bm A_n)({\cal D}^n\bm A^m)
    -{ig}\epsilon^{mn} \bm A_m\ad{\bm f}\bm A_n
    \right],\\\label{quad13}
  {\cal L}^{(2)}_3
  &= 2\Tr[({\cal D}^m\bm A^\mu)(\der_\mu \bm A_m)], 
\end{align}
where ${\cal L}^{(2)}_1$ and ${\cal L}^{(2)}_2$ are quadratic terms
for $\bm A_\mu$ and $\bm A_m$, respectively. There appears a mixing
term between $\bm A_\mu$ and $\bm A_m$ in ${\cal L}^{(2)}_3$. As seen
below, this mixing term can be canceled by gauge fixing terms.

Using integration by parts, we can rewrite the quadratic terms in
eqs.~\eqref{quad11}--\eqref{quad13} as
\begin{align}\label{L1201}
  {\cal L}^{(2)}_1
  &=\Tr[
\bm A^\mu(\eta_{\mu\nu}\square +\eta_{\mu\nu} {\cal D}_m{\cal D}^m-\der_\mu\der_\nu)
\bm A^\nu],\\
  {\cal L}^{(2)}_2
  &=\Tr[
    \bm A^m(\delta_{mn}\square +\delta_{mn} {\cal D}_{m'}{\cal D}^{m'}-
    {\cal D}_n{\cal D}_m-ig\epsilon_{mn}\ad{\bm f}  )
    \bm A^n],\\
  {\cal L}^{(2)}_3
  &= 2\Tr[(\der_\mu\bm A^\mu)({\cal D}_m \bm A^m)],   
    \label{L1203}
\end{align}
where $\square=\der_\mu\der^\mu$ is the 4D d'Alembertian. In the above
expressions, we discarded the surface terms. Since we consider the
compact extra dimensions, the surface terms have to be carefully
treated. In the present case, all of the surface terms actually vanish
in the action as discussed in appendix~\ref{app:surf}.

\subsection{Gauge fixing}
\label{sec:gaugefix}
We consider the standard procedure to fix the gauge in quantum
theories. The gauge fixing yields the additional contributions to the
Lagrangian as
\begin{align}\label{gflag1}
  {\cal L}_{\rm YM}^{\rm gf}= {\cal L}_{\rm YM}+{\cal L}_{\rm GF}+{\cal L}_{c}.
\end{align}
We refer to ${\cal L}_{\rm GF}$ and ${\cal L}_{c}$ as gauge fixing
terms and ghost terms, respectively. We choose the gauge fixing terms
as
\begin{align}\label{gflag2}
  {\cal L}_{\rm GF}
  &=-{1\over \xi}\Tr[(\der_\mu \bm A^\mu+\xi {\cal D}_m\bm A^m)^2]\\
  &={1\over \xi}\Tr[\bm A^\mu\der_\mu \der_\nu \bm A^\nu]+\xi \Tr[\bm A^m{\cal D}_m{\cal D}_n\bm A^n]-2\Tr[(\der_\mu\bm A^\mu)({\cal D}_m\bm A^m)],
        \label{gflag3}
\end{align}
where $\xi\in \mathbb R$ is called the gauge fixing parameter. From
the first line to the second line, we have used integration by parts
and dropped the surface terms, which vanish in the action as shown in
appendix~\ref{app:surf}. One sees that the last term in
eq.~\eqref{gflag3} cancels ${\cal L}^{(2)}_3$ in
${\cal L}_{\rm YM}^{\rm gf}$.

From the gauge fixing terms given above, the ghost terms are obtained as
\begin{align}
  {\cal L}_{c}
  & =-2\Tr[\bar {\bm c}\left\{\der^\mu(\der_\mu-ig\ad{\bm A_\mu})+\xi {\cal D}^m
    ({\cal D}_m-ig \ad{\bm A_m}) \right\} \bm c],
\end{align}
where $\bm c$ and $\bar {\bm c}$ are ghost fields. They can be
expanded by the $su(n)$ generators.  In addition to the quadratic
terms for the ghost fields, there also appear trilinear interactions
including the gauge field fluctuations.

\subsection{Quadratic terms}
\label{sec:quad}
To discuss the tree-level mass spectrum, let us focus on the quadratic
terms of the fluctuations $\bm A_M$ and ghost fields. From the
discussions in section~\ref{sec:fluc} and~\ref{sec:gaugefix}, these
terms in the Lagrangian in eq.~\eqref{gflag1} are given by
\begin{align}
  {\cal L}_{\rm YM}^{\rm gf}
  &\ni {\cal L}^{(2)}_{A_\mu}
  +{\cal L}^{(2)}_{A_m}
    +{\cal L}^{(2)}_{c}, \\
  {\cal L}^{(2)}_{A_\mu}
  &=\Tr[
    \bm A^\mu(\eta_{\mu\nu}\square
    +\eta_{\mu\nu} {\cal D}_m{\cal D}^m-(1-\xi^{-1})\der_\mu\der_\nu)
    \bm A^\nu],\\
  {\cal L}^{(2)}_{A_m}
  &=\Tr[
    \bm A^m(\delta_{mn}\square +\delta_{mn} {\cal D}_{m'}{\cal D}^{m'}
    -(1-\xi) {\cal D}_m{\cal D}_n
    -2ig\epsilon_{mn}\ad{\bm f}  )
    \bm A^n],
    \label{Lam2Tr1}
    \\
  {\cal L}^{(2)}_{c}
  &=-2\Tr[\bar {\bm c}(\square+\xi {\cal D}_m
    {\cal D}^m)\bm c].
\end{align}
We note that, to obtain the expression for ${\cal L}^{(2)}_{A_m}$, we have used
\begin{align}
  {\cal D}_n {\cal D}_m
  &=[{\cal D}_n ,{\cal D}_m ]+{\cal D}_m{\cal D}_n
    =ig \epsilon_{mn}\ad{\bm f}+{\cal D}_m{\cal D}_n.
\end{align}
The last term in eq.~\eqref{Lam2Tr1} being proportional to
$\ad{\bm f}$ appears since the extra-dimensional components of the
gauge fields have non-trivial helicities in the two-dimensional
torus~\cite{Bachas}.

We use the explicit forms of the generators in
eqs.~\eqref{Hnormfund1},~\eqref{Hnormfund2}, and~\eqref{Eijdef1} to
expand $\bm A_M$ as
\begin{align}\label{expandAM1}
  \bm A_M=A_M^k\hat H_k+
  A_M^{ij}E_{ij}^{(+)}+  \bar A_M^{ij}E_{ij}^{(-)}
=
  \begin{pmatrix}
      A_M^1 & A_M^{12} & \dots & A_M^{1n}\\
      \bar A_M^{12} & A_M^2-A_M^1& &\vdots
      \\
      \vdots & & \ddots & \\
      \bar A_M^{1n} & \dots 
      & & -A_M^{n-1}
  \end{pmatrix},
\end{align}
where we have introduced the component fields $A_M^k$
$(k=1,\dots,n-1)$, $A_M^{ij}$, and $\bar A_M^{ij}=(A_M^{ij})^\dag$
$(1\leq i<j\leq n)$. We note that the summations over
$1\leq k\leq n-1$ or $1\leq i<j\leq n$ are implied in each term in the
second expression in eq.~\eqref{expandAM1}. A similar expansion to
eq.~\eqref{expandAM1} for the ghost fields is also used.

The expansion in eq.~\eqref{expandAM1} simplifies the discussion since
the component fields are eigenfunctions of the background covariant
derivative and the boundary conditions. We first express the
Lagrangian by the component fields.  From eq.~\eqref{fecom1}, one sees
that
\begin{align}
  {\cal D}_m\bm A_M=\der_m
  A_M^k\hat H_k+
D_m^{(ij)}  A_M^{ij}E_{ij}^{(+)}+ \bar D_m^{(ij)} \bar A_M^{ij}E_{ij}^{(-)}, 
\end{align}
where we have defined
\begin{align}\label{covariantderivative}
& D_5^{(ij)}=\der_5+ig\hat f \tilde N^{ij}(1+\gamma)x^6/2, \qquad 
  D_6^{(ij)}=\der_6-ig\hat f \tilde N^{ij}(1-\gamma)x^5/2, \\
& \bar D_5^{(ij)}=\der_5-ig\hat f \tilde N^{ij}(1+\gamma)x^6/2, \qquad 
  \bar D_6^{(ij)}=\der_6+ig\hat f \tilde N^{ij}(1-\gamma)x^5/2.
\end{align}
Thus, each component is an eigenfunction of the covariant
derivative. Using eqs.~\eqref{trgen1} and~\eqref{trgen2}, we obtain
\begin{align}\label{L2CWb1}
  {\cal L}^{(2)}_{A_\mu}
  &=(M^{su(n)})_{k\ell}A_\mu^{k}\left[\eta^{\mu\nu}\square
    -(1-\xi^{-1})\der^\mu\der^\nu+\eta^{\mu\nu} \der_m\der^m\right]A_\nu^{\ell}
    +\sum_{1\leq i<j\leq n}{\cal L}^{(ij)}_{A_\mu}, \\\label{L2CWb2}
  {\cal L}^{(2)}_{A_m}
  &=(M^{su(n)})_{k\ell}A_m^{k}\left[\eta^{mn}\square
    +\eta^{mn} \der_{m'}\der^{m'}-(1-\xi)\der^m\der^n\right]A_n^{\ell}
    +\sum_{1\leq i<j\leq n}{\cal L}^{(ij)}_{A_m},
\end{align}
where we have introduced 
\begin{align}\label{qlagAmu1}
  {\cal L}^{(ij)}_{A_\mu}
  &=2\bar A_\mu^{ij}[\eta^{\mu\nu}\square-(1-\xi^{-1})\der^\mu\der^\nu +\eta^{\mu\nu} (D_{m'}^{(ij)})^2 ] A_\nu^{ij}, \\
  {\cal L}^{(ij)}_{A_m}
  &=2\bar A_m^{ij}[\delta^{mn}\square +\delta^{mn} (D_{m'}^{(ij)})^2
    -(1-\xi) D^{m(ij)}D^{n(ij)}
    -2ig\hat f \tilde N^{ij}\epsilon^{mn}  ] A_n^{ij}    .
    \label{qlagAm1}
\end{align}
For the ghost fields, a similar discussion holds.  To obtain
eqs.~\eqref{qlagAmu1} and~\eqref{qlagAm1}, we have used integration by
parts and safely discarded surface terms; please refer to
appendix~\ref{app:surf}.

In the Lagrangian in eqs.~\eqref{L2CWb1} and~\eqref{L2CWb2}, the
components related to the Cartan generators $\hat H_k$ do not couple
to flux. Although the Cartan matrix $M^{su(n)}$ is not diagonalized in
this basis, we can take linear combinations of $A_M^{k}$ to
diagonalize the Lagrangian.  For the components related to the step
operators, the Lagrangian is completely separated for each $(ij)$.

Let us discuss the boundary conditions for the component fields.  As
discussed in section~\ref{sec:bcT2}, the fluctuations $\bm A_M$
satisfy the boundary conditions in eq.~\eqref{bcfluc1}. Using the
expression of the twist matrices in eq.~\eqref{bmnewg2}, we can
rewrite the boundary conditions in eq.~\eqref{bcfluc1} as
\begin{align}
  \bm A_M({\cal T}_5z)
  &=e^{-ig\ad{\bm v_5}}e^{ig(1-\gamma)(x^6/2)\ad{\bm f}}\bm A_M(z), \\
  \bm A_M({\cal T}_6z)
  &=e^{-ig(\tau_{\rm R}\ad{\bm v_5}+\tau_{\rm I}\ad{\bm v_6})}
    e^{ig\{-(1+\gamma) \tau_{\rm I}x^5/2+(1-\gamma) \tau_{\rm R}x^6/2\}\ad{\bm f}}
    \bm A_M(z) , 
\end{align}
where we have used the Campbell-Baker-Hausdorff formula
$e^XYe^{-X}=e^{\ad{X}}Y$. Combining the above and
eq.~\eqref{expandAM1}, one sees that the components obey the following
boundary conditions:
\begin{align}\label{bcCartanA1}
  A_M^k({\cal T}_5z)
  &=A_M^k({\cal T}_6z)=A_M^k(z), \\\label{bcAijT5}
  A_M^{ij}({\cal T}_5z)
  &=e^{-ig\tilde v_5^{ij}}e^{ig(1-\gamma)(x^6/2)\hat f \tilde N^{ij}}A_M^{ij}(z), \\
  \label{bcAijT6}
  A_M^{ij}({\cal T}_6z)
  &=
    e^{-ig(\tau_{\rm R}\tilde v_5^{ij}+\tau_{\rm I}\tilde v_6^{ij})}
    e^{ig\{-(1+\gamma) \tau_{\rm I}x^5/2+(1-\gamma) \tau_{\rm R}x^6/2\}\hat f\tilde N^{ij}}A_M^{ij}(z), 
\end{align}
where the ones for $\bar A_M^{ij}$ are given by $\bar A_M^{ij}=(A_M^{ij})^\dag$.

From the Lagrangian and the boundary conditions for the component
fields, we can understand that the components $A_M^k$ are decoupled
from both the flux and the continuous Wilson lines. On the other hand,
the components $A_M^{ij}$ can couple with them if $\tilde N^{ij}$ or
$\tilde v_m^{ij}$ do not vanish. We keep general configurations of the
Wilson line phases. Then, for a given magnetic flux configuration,
i.e., a value of $N^k$ in eq.~\eqref{qcondhatb1}, the masses for
$A_M^{ij}$ are basically classified into $\tilde N^{ij}=0$ or
$\tilde N^{ij}\neq 0$ cases. We will examine the mass spectrum for
both cases in the next section.

\section{Mass spectrum around flux backgrounds}
\label{sec:treemass}

\subsection{Masses for KK modes of $A_\mu^k$}
\label{sec:kkmassAmuk}
Let us discuss the tree-level mass spectrum obtained from
$A_\mu^k$. Their masses are derived from the 6D kinetic terms in
eq.~\eqref{L2CWb1} and the boundary conditions in
eq.~\eqref{bcCartanA1}. As they are periodic under the translations
generated by ${\cal T}_p$, it is useful to introduce Kaluza-Klein (KK)
expansions as
\begin{align}\label{amuKK1}
  A_\mu^k(x^\mu,z)=\sum_{\hat n,\hat m\in \mathbb Z}C^{(\hat n,\hat m)}e^{2\pi i\hat n x^5}
  e^{2\pi i \tau_{\rm I}^{-1}(\hat m-\hat n\tau_{\rm R})x^6}A_{\mu(\hat n,\hat m)}^{k}(x^\mu), 
\end{align}
where $C^{(\hat n,\hat m)}$ is a normalization constant, and
$A_{\mu(\hat n,\hat m)}^{k}(x^\mu)$ $(\hat n,\hat m\in \mathbb Z)$ is
identified as a 4D field.

We denote the KK mass for $A_{\mu(\hat n,\hat m)}^{k}$ as
$M^2(A_{\mu(\hat n,\hat m)}^{k})$.  The Lagrangian in
eq.~\eqref{L2CWb1} implies that $M^2(A_{\mu(\hat n,\hat m)}^{k})$ is
given by the eigenvalue of the operator $-\der_m\der^m$ acting on the
corresponding mode function. Thus, we obtain
\begin{align}\label{kkamuk}
  M^2(A_{\mu(\hat n,\hat m)}^{k})=\left({2\pi\over L}\right)^2\left[\hat n^2+
  \tau_{\rm I}^{-2}(\hat m
  -\hat n\tau_{\rm R})^2\right],
\end{align}
where we have temporarily written $L$ for an illustration.

The mass spectrum contains the massless 4D gauge fields
$A_\mu^{k(0,0)}$, namely, the zero modes. The massless gauge fields
are related to a gauge symmetry manifested in a low-energy effective
theory.

There is a clear geometrical interpretation of the KK masses in
eq.~\eqref{kkamuk}. The torus lattice is spanned by the basis vectors
$\bm e_1=(1,0)$ and $\bm e_2=(\tau_{\rm R},\tau_{\rm I})$, where the
right-hand sides are components in an orthogonal basis. The basis
vectors $\bm e^i$ of a dual lattice are defined by
$\bm e_i\cdot \bm e^j=2\pi \delta_i^j$, which gives
$\bm e^1=2\pi (1,-\tau_{\rm R}/\tau_{\rm I})$ and
$\bm e^2=2\pi (0,1/\tau_{\rm I})$.  Discretized extra-dimensional
momenta correspond to points on the dual lattice, and the squared norm
of a dual vector from the origin to a point,
\begin{align}
  |\hat n\bm e^1+\hat m\bm e^2|^2
  &=(2\pi)^2\left[\hat n^2+\tau_{\rm I}^{-2}(\hat m-\hat n\tau_{\rm R})^2\right], 
\end{align}
gives the KK mass squared in eq.~\eqref{kkamuk}.

\subsection{Masses for KK modes of $A_m^k$}
\label{sec:massAmk}
Let us study the mass spectrum obtained from $A_m^k$. Since $A_m^k$
obeys the same boundary conditions as those for $A_\mu^k$, the KK
expansion of $A_m^k$ is given by using the same mode functions as in
eq.~\eqref{amuKK1}. We write the KK mode of $A_m^k$ as
$A_{m(\hat n,\hat m)}^{k}$.

The KK masses for $A_{m(\hat n,\hat m)}^{k}$, denoted by
$M^2(A_{m(\hat n,\hat m)}^{k})$, are determined by the Lagrangian in
eq.~\eqref{L2CWb2}. The mass matrices for
$(A_{5(\hat n,\hat m)}^{k},A_{6(\hat n,\hat m)}^{k})$ appear from the
operator $-\eta^{mn} \der_{m'}\der^{m'}+(1-\xi)\der^m\der^n$ in
eq.~\eqref{L2CWb2} and are given by
\begin{align}\notag
  &  (A_{5(\hat n,\hat m)}^{k}
    ~A_{6(\hat n,\hat m)}^{k})
  \left({2\pi\over L}\right)^2\bigg[
  \left[\hat n^2+\tau_{\rm I}^{-2}(\hat m
  -\hat n\tau_{\rm R})^2\right]
  \begin{pmatrix}
      1&0\\0&1
  \end{pmatrix}
  \\
&\qquad\qquad\qquad\qquad\qquad -(1-\xi)
       \begin{pmatrix}
           \hat n^2 &\hat n\tau_{\rm I}^{-1}(\hat m
           -\hat n\tau_{\rm R})\\
           \hat n\tau_{\rm I}^{-1}(\hat m
  -\hat n\tau_{\rm R}) &\tau_{\rm I}^{-2}(\hat m
  -\hat n\tau_{\rm R})^2
       \end{pmatrix}
  \bigg]
                         \begin{pmatrix}
                             A_{5(\hat n,\hat m)}^{k}\\
                             A_{6(\hat n,\hat m)}^{k}                            
                         \end{pmatrix}. 
\end{align}
Diagonalizing the above, we find mass eigenvalues
$M_{\rm ph}^2(A_{m(\hat n,\hat m)}^{k})$ and
$M_{\xi}^2(A_{m(\hat n,\hat m)}^{k})$ for each $(\hat n,\hat m)$ mode
as
\begin{align}\label{kkamk1}
  M_{\rm ph}^2(A_{m(\hat n,\hat m)}^{k})
  &=\left({2\pi\over L}\right)^2\left[\hat n^2+\tau_{\rm I}^{-2}(\hat m
  -\hat n\tau_{\rm R})^2\right], \\\label{kkamk2}
  M_\xi^2(A_{m(\hat n,\hat m)}^{k})
  &=  \xi \left({2\pi\over L}\right)^2\left[\hat n^2+\tau_{\rm I}^{-2}(\hat m
  -\hat n\tau_{\rm R})^2\right].
\end{align}

The zero modes $A_{m(0,0)}^k$ are massless scalars whose masses are
independent of the gauge parameter $\xi$.  Some of these zero modes
may acquire non-zero masses through quantum corrections, except for NG
bosons associated with the breaking of the translational invariance on
$T^2$ via non-vanishing flux~\cite{Buchmuller,B1}.  We will discuss
the quantum corrections in section~\ref{subsec:potNG}.  For
$(\hat n,\hat m)\neq (0,0)$, there appear massive scalar modes. The
massive scalars with masses $M_{\rm ph}^2(A_{m(\hat n,\hat m)}^{k})$
are degenerate with $A_{\mu(\hat n,\hat m)}^{k}$, although the massive
modes with masses $M_\xi^2(A_{m(\hat n,\hat m)}^{k})$ are would-be
Goldstone modes, which provide physical degrees of freedom to
longitudinal modes of massive vector fields
$A_{\mu(\hat n,\hat m)}^{k}$.

\subsection{Masses for KK modes of $A_\mu^{ij}$ and $A_m^{ij}$ with $\tilde N^{ij}=0$}
\label{sec:}
We discuss the mass spectrum that arises from $A_\mu^{ij}$ and
$A_m^{ij}$ in the $\tilde N^{ij}=0$ case. In this case, the
Lagrangians in eqs.~\eqref{qlagAmu1} and~\eqref{qlagAm1} are
simplified to
\begin{align}\label{qlagAmu1Nzero}
  {\cal L}^{(ij)}_{A_\mu}
  &\to 2\bar A_\mu^{ij}[\eta^{\mu\nu}\square-(1-\xi^{-1})\der^\mu\der^\nu +\eta^{\mu\nu} \der_{m'}\der^{m'} ] A_\nu^{ij}, \\
  {\cal L}^{(ij)}_{A_m}
  &\to 2\bar A_m^{ij}[\delta^{mn}\square +\delta^{mn} \der_{m'}\der^{m'}
    -(1-\xi) \der^m\der^n ] A_n^{ij}    .
\end{align}
The differential operators appearing above are the same ones as in the
Lagrangian for $A_M^{k}$. On the other hand, the boundary conditions
for $A_M^{ij}$ are different from those for $A_M^k$.  From
eqs.~\eqref{bcAijT5} and~\eqref{bcAijT6}, one sees that $A_M^{ij}$ are
not periodic under the translations for $\tilde v_m^{ij}\neq 0$, while
$A_M^k$ are periodic.

Due to the phase factors including $\tilde v_m^{ij}$ in
eqs.~\eqref{bcAijT5} and~\eqref{bcAijT6}, the mass spectrum is
modified compared to that of $A_M^k$. These phase factors induce the
overall shifts of the momentum lattice spanned by the dual basis
vectors $\bm e^i$ discussed in section~\ref{sec:kkmassAmuk}.  As a
result, KK masses for $A_\mu^{ij}$ and $A_m^{ij}$ with
$\tilde N^{ij}=0$ have similar forms to the masses in
eqs.~\eqref{kkamuk}, \eqref{kkamk1}, and \eqref{kkamk2}, but $\hat n$
and $\hat m$ are replaced by $\hat n-g\tilde v_5^{ij}/2\pi$ and
$\hat m-g\tilde v_6^{ij}/2\pi$, respectively.\footnote{A similar
  argument in a $T^2/\mathbb Z_3$ orbifold model, for instance, is
  found in~\cite{Kojima:2023mew}.} Let us introduce the
parametrization $\tilde a_m^{ij}=g\tilde v_m^{ij}/2\pi$. Then, the KK
masses for $A_\mu^{ij}$ and $A_m^{ij}$ are given by
\begin{align}
  M^2(A_{\mu(\hat n,\hat m)}^{ij})&=
  M_{\rm ph}^2(A_{m(\hat n,\hat m)}^{ij})\\
  &=
  \left({2\pi\over L}\right)^2\left[(\hat n-\tilde a_5^{ij})^2+\tau_{\rm I}^{-2}((\hat m-\tilde a_6^{ij})
  -(\hat n-\tilde a_5^{ij})\tau_{\rm R})^2\right],\\
  M_\xi^2(A^{ij}_{m(\hat n,\hat m)})
  &=
  \xi\left({2\pi\over L}\right)^2\left[(\hat n-\tilde a_5^{ij})^2+\tau_{\rm I}^{-2}((\hat m-\tilde a_6^{ij})
  -(\hat n-\tilde a_5^{ij})\tau_{\rm R})^2\right].
\end{align}
Note that the KK mass spectrum is invariant under integer shifts of
$\tilde a^{ij}_m$.  This property is expected from the boundary
conditions in eqs.~\eqref{bcAijT5} and~\eqref{bcAijT6}, which are also
invariant under the integer shifts.

Except for the case with $\tilde a_m^{ij}=0$ mod~1, this spectrum has
no massless modes.  As in the case of $A_M^k$, massive 4D vector
fields from $A_\mu^{ij}$ and scalar fields from $A_m^{ij}$ appear.
Half of the scalar KK modes have $\xi$-dependent masses
$M_\xi^2(A_{m(\hat n,\hat m)}^{ij})$ and are would-be Goldstone modes
that provide physical degrees of freedom to the longitudinal modes of
massive vector fields.

\subsection{Masses for Landau level excitations of $A_\mu^{ij}$ with
  $\tilde N^{ij}\neq 0$}
\label{sec:Aijlandaumass}
We discuss the mass spectrum of the gauge field $A_\mu^{ij}$ in the
$\tilde{N}^{ij}\neq 0$ case.  In this case, $A_\mu^{ij}$ couples to
the background flux through the covariant derivative in
eq.~\eqref{qlagAmu1}.  The mass spectrum is determined by the
eigenvalues of the operator
\begin{align}\label{massOPL1}
  -(D_m^{(ij)})^2=[-i\der_5+g\hat f \tilde N^{ij}(1+\gamma)x^6/2]^2
  +[-i\der_6-g\hat f \tilde N^{ij}(1-\gamma)x^5/2]^2. 
\end{align}
The mode function is given as the eigenfunction of this operator and should be consistent with the boundary conditions in eqs.~\eqref{bcAijT5} and~\eqref{bcAijT6}.

This system is an analog to a two-dimensional quantum mechanical
system with a constant magnetic flux. The mass spectrum is the
well-known Landau levels~\cite{T1}. Let us denote the quantum
mechanical momentum operator $p_m=-i\der_m$, which satisfies
$[x^m,p_n]=i\delta_n^m$. It is convenient to express
\begin{align}
  -iD_5^{(ij)}=p_5+g\hat f \tilde N^{ij}(1+\gamma)x^6/2, \qquad
  -iD_6^{(ij)}=p_6-g\hat f \tilde N^{ij}(1-\gamma)x^5/2,
\end{align}
which satisfy $[-iD_5^{(ij)},-iD_6^{(ij)}]=ig\hat f\tilde N^{ij}$
independently to $\gamma$.  Let us define
\begin{align}\label{Pi} 
  \Pi_z^{(ij)}
  &=-i\sqrt{2\over g\hat f
    |\tilde N^{ij}|}{D_5^{(ij)}-iD_6^{(ij)}\over 2}
    =\sqrt{2\over g\hat f |\tilde N^{ij}|}
    \left[p_z+i{g\hat f \tilde N^{ij}\over 4}(\bar z-\gamma z)\right], \\
    \label{Pibar}
  \bar \Pi_z^{(ij)}
  &=-i
    \sqrt{2\over g\hat f |\tilde N^{ij}|}{D_5^{(ij)}+iD_6^{(ij)}\over 2}
  =\sqrt{2\over g\hat f |\tilde N^{ij}|}\left[\bar p_z-i{g\hat f \tilde N^{ij}\over 4}(z-\gamma \bar z)\right], 
\end{align}
where we have used $p_z=-i\der_z$ and $\bar p_z=-i\bar \der_z$.

For the case with $\tilde N^{ij}>0$ $(\tilde N^{ij}<0)$,
$[\bar \Pi_z^{(ij)},\Pi_z^{(ij)}]=1$
($[\Pi_z^{(ij)},\bar \Pi_z^{(ij)}]=1$) holds.  Thus,
$\bar \Pi_z^{(ij)}$ and $\Pi_z^{(ij)}$ ($\Pi_z^{(ij)}$ and
$\bar \Pi_z^{(ij)}$) are interpreted as an annihilation and a creation
operator, respectively. The operator in eq.~\eqref{massOPL1} is
expressed by them as
\begin{align}
-(D_m^{(ij)})^2=2g\hat f|\tilde N^{ij}| (\hat a^\dag \hat a+1/2),
\label{dmsqNnex01}
\end{align}
where $(\hat a,\hat a^\dag)=(\bar \Pi_z^{(ij)},\Pi_z^{(ij)})$ for
$\tilde N^{ij}>0$, and $(\hat a,\hat a^\dag)=(\Pi_z^{(ij)},\bar \Pi_z^{(ij)})$
for $\tilde N^{ij}<0$.

Eigenvalues of the operator in eq.~\eqref{dmsqNnex01} are the same as
those of a harmonic oscillator. Also, as shown in
appendix~\ref{appendix:modefunction}, the eigenvalues are derived by
introducing the mode functions of the ground state
$\zeta_{0,d}^{ij}(z)$ and $\ell$--th excited states
$\zeta_{\ell,d}^{ij}(z)$ $(\hat \ell\in \mathbb Z_{\geq 1})$ as
\begin{align}\label{modefunction}
  \hat a \zeta_{0,d}^{ij}(z)=0, \qquad
  \zeta_{\hat \ell,d}^{ij}(z) ={1\over \sqrt{\hat \ell!}}(\hat a^\dag)^{\hat \ell}\zeta^{ij}_{0,d}(z).
\end{align}
The subscript $d$ takes $d=0,\dots, |\tilde N^{ij}|-1$ and labels
$|\tilde N^{ij}|$ degenerate states in this system. We also have to
impose that the mode functions obey the same boundary conditions for
$A_M^{ij}$ in eqs.~\eqref{bcAijT5} and~\eqref{bcAijT6} as
\begin{align}
    \label{bczetaijT5}
    \zeta^{ij}_{\hat \ell,d}({\cal T}_5z)
  &=e^{-2\pi i\tilde a_5^{ij}}e^{ig(1-\gamma)(x^6/2)\hat f \tilde N^{ij}}\zeta^{ij}_{\hat \ell,d}(z), \\
  \label{bczetaijT6}
  \zeta^{ij}_{\hat \ell,d}({\cal T}_6z)
  &=
    e^{-2\pi i(\tau_{\rm R}\tilde a_5^{ij}+\tau_{\rm I}\tilde a_6^{ij})}
    e^{ig\{-(1+\gamma) \tau_{\rm I}x^5/2+(1-\gamma) \tau_{\rm R}x^6/2\}\hat f\tilde N^{ij}}\zeta^{ij}_{\hat \ell,d}(z). 
\end{align}
The explicit form of the mode function is given by
\begin{align}
    \label{eqn:zetapls}
    \zeta_{\hat \ell,d}^{ij}(z)
    =&
    \left(\frac{2\pi \tilde{N}^{ij}}{\tau_{\rm I}} \right)^{1/4}
    \frac{1}{2^{\hat \ell} \sqrt{\hat \ell !}}
    e^{\frac{\pi \tilde{N}^{ij}}{2\tau_{\rm I}}[z(z-\bar{z})-\frac{\gamma}{2}(z+\bar{z})(z-\bar{z})]}
    \\
    &\times
    \sum_{n=-\infty}^{+\infty}
    H_{\hat \ell}(w_{n,d}(z))
    e^{i\pi\tilde{N}^{ij}\tau (n- (\tilde a_5^{ij}-d)/\tilde{N}^{ij})^2}
        e^{2\pi i(n- (\tilde a_5^{ij}-d)/\tilde{N}^{ij})(\tilde{N}^{ij}z+\tau_{\rm R}\tilde a_5^{ij}+\tau_{\rm I}\tilde a_6^{ij}-\gamma \tilde{N}^{ij}\tau_{\rm R}/2)},
        \notag
\end{align}
for $\tilde{N}^{ij}>0$, where $H_{\hat \ell} (x)$ are the Hermite
polynomials, and we have used
\begin{align}
    w_{n,d}(z)
    =
    \sqrt{\frac{2\pi N^{ij}}{\tau_{\rm I}}}
    \left[
    \frac{z-\bar{z}}{2i}
    +\tau_{\rm I}\left(n-\frac{\tilde a_5^{ij}-d}{\tilde{N}^{ij}} \right)
    \right].
\end{align}
The normalization constant has been determined so that the orthogonal
relation is satisfied as
\begin{align}\label{orthogonal}
    \int_{\mathcal{V}_{T^2}} dx^5dx^6\
    \bar{\zeta}_{\hat \ell,d}^{ij}
    \zeta_{\hat \ell',d'}^{ij}
    =
    \delta_{\hat \ell \hat \ell'}\delta_{dd'},
\end{align}
where
$\bar{\zeta}_{\hat \ell,d}^{ij}=({\zeta}_{\hat \ell,d}^{ij})^\dag$.
One can obtain the explicit mode function for $\tilde{N}^{ij}<0$ by
taking the complex conjugate of eq.~\eqref{eqn:zetapls}.  The
derivation and details are summarized in
appendix~\ref{appendix:modefunction}.  Through the boundary conditions
in eqs.~\eqref{bczetaijT5} and~\eqref{bczetaijT6}, the mode function
in eq.~\eqref{eqn:zetapls} depends on the Wilson line phases
$\tilde a_m^{ij}$.  As explained above, there are $\tilde N^{ij}$
independent mode functions for a fixed $\hat \ell$.  From the
right-hand side of eq.~\eqref{eqn:zetapls}, one sees that the shift of
$d\to d + \tilde N^{ij}$ leaves the mode function unchanged.

Let us define the mode expansion as
\begin{align}\label{amuijexp1}
  A_\mu^{ij}(x^\mu ,z)=\sum_{\hat \ell=0}^\infty \sum_{d=1}^{|\tilde N^{ij}|}
  A_{\mu(\hat \ell,d)}^{ij}(x^\mu)  \zeta^{ij}_{\hat \ell,d}(z), 
\end{align}
where we refer to $A_{\mu(\hat \ell,d)}^{ij}(x^\mu)$ as the
$\hat \ell$--th Landau level. From the operator in
eq.~\eqref{massOPL1} and the mode expansion in eq.~\eqref{amuijexp1},
we find that the masses for the $\hat \ell$--th Landau level, denoted
by $M^2(A_{\mu(\hat \ell,d)}^{ij})$, are given by
\begin{align}
  M^2(A_{\mu(\hat \ell,d)}^{ij})
  =2g\hat f|\tilde N^{ij}| (\hat \ell+1/2).
\end{align}
There are no massless modes, and all $A_{\mu(\hat \ell,d)}^{ij}$ have
masses proportional to $L^{-1}$. Thus, the gauge symmetry is broken by
the flux.  We note that Wilson line phases do not appear in the mass
spectrum, although the mode function in eq.~\eqref{eqn:zetapls}
depends on the phases.

\subsection{Masses for Landau level excitations of $A_m^{ij}$ with
  $\tilde N^{ij}\neq 0$}
\label{sec:ammasswt}
In this section, we discuss the mass spectrum for $A_m^{ij}$ when
$\tilde N^{ij}\neq 0$, that is, the components that couple to the
flux.  The mass spectrum is determined by the last three terms in the
Lagrangian in eq.~\eqref{qlagAm1}, where $A_5^{ij}$ and $A_6^{ij}$ are
mixed.  After performing the diagonalization of this term, it is
convenient to define
\begin{align}
  \begin{pmatrix}
    A_-^{ij}\\A_+^{ij}
  \end{pmatrix}
  ={1\over \sqrt{2}}
  \begin{pmatrix}
    A_5^{ij}-iA_6^{ij}\\
    A_5^{ij}+iA_6^{ij}
  \end{pmatrix},
\end{align}
where $A_-^{ij}$ and $A_+^{ij}$ correspond to components of $\bm A_z$
and $\bar {\bm A}_z$ in eq.~\eqref{defaz1} related to the generator
$E_{ij}^{(+)}$ in eq.~\eqref{Eijdef1} and
$\bar A_\pm^{ij}=(A_\pm^{ij})^\dag$.  With this definition, the
Lagrangian in eq.~\eqref{qlagAm1} is rewritten as
\begin{align}\notag
    {\cal L}^{(ij)}_{A_m} &= 2\bar A_-^{ij} \left[\square + \frac{1+\xi}{2} (D_{m}^{(ij)})^2 +\frac{3+\xi}{2} g\hat f \tilde N^{ij} \right] A_-^{ij}
    \\\notag
    &\quad+2\bar A_+^{ij}\left[
    \square + \frac{1+\xi}{2} (D_{m}^{(ij)})^2 -\frac{3+\xi}{2} g\hat f \tilde N^{ij}
    \right] A_+^{ij}
\\
    &\quad-2
    \bar A_+^{ij}\left[
    \frac{1-\xi}{2}(D_{5}^{(ij)}+iD_{6}^{(ij)})^2
    \right] A_-^{ij}-2
    \bar A_-^{ij}\left[
    \frac{1-\xi}{2}(D_{5}^{(ij)}-iD_{6}^{(ij)})^2
    \right] A_+^{ij}, 
\end{align}
where the covariant derivatives are defined in
eq.~\eqref{covariantderivative}.  From the definitions of
$\Pi_z^{(ij)}$ and $\bar \Pi_z^{(ij)}$ in eqs.~\eqref{Pi}
and~\eqref{Pibar}, respectively, the Lagrangian can be written in
terms of creation and annihilation operators
\begin{equation}
    {\cal L}^{(ij)}_{A_m} =2
  \begin{pmatrix}
      \bar A_-^{ij}
      &\bar A_+^{ij}
  \end{pmatrix}
  \begin{pmatrix}
      \square-g\hat f\tilde N^{ij}((1+\xi)\hat a^\dag \hat a-1)
      &
      g\hat f\tilde N^{ij}(1-\xi)\hat a^\dag\hat a^\dag\\
      g\hat f\tilde N^{ij}(1-\xi)\hat a\hat a
      &\square-g\hat f\tilde N^{ij}((1+\xi)\hat a^\dag \hat a+\xi+2)      
  \end{pmatrix}
        \begin{pmatrix}
            A_-^{ij}\\ A_+^{ij}
        \end{pmatrix},
\end{equation}
where we have set $\tilde N^{ij}>0$ for simplicity. In this case,
$(\hat a,\hat a^\dag)=(\bar \Pi_z^{(ij)},\Pi_z^{(ij)})$.  For the
$\tilde N^{ij}<0$ case, we also have a similar expression.

Since $A_m^{ij}$ obeys the same boundary conditions as $A_{\mu}^{ij}$,
they have the same mode functions in the KK expansion in
eq.~\eqref{amuijexp1}.  Therefore, the mode expansion is defined as
\begin{align}
  A_\pm^{ij}(x^\mu,z)=\sum_{\hat \ell=0}^\infty \sum_{d=1}^{\tilde N^{ij}}
  A_{\pm(\hat \ell,d)}^{ij}(x^\mu)\zeta^{ij}_{\hat \ell,d}(z), 
\end{align}
where $A_{\pm(\hat \ell,d)}^{ij}$ is the $\hat \ell$--th Landau level,
and $\zeta_{\hat \ell,d}^{ij}(z)$ satisfy orthonormal relations as
already mentioned in eq.~\eqref{orthogonal}.  After acting the
creation and annihilation operators on the mode expansion and
integrating out the extra dimensions, the Lagrangian becomes
\begin{align}\label{lag4Dam3}
 {\cal L}^{(ij)}_{A_m} &= 2\bar A_{-(0,d)}^{ij} \left[\square+g\hat f \tilde N^{ij} \right] A_{-(0,d)}^{ij}
    +2\bar A_{-(1,d)}^{ij} \left[\square-\xi g\hat f \tilde N^{ij} \right] A_{-(1,d)}^{ij}
    \\\notag
    &+2\sum_{\hat \ell=0}^\infty 
  \begin{pmatrix}
      \bar A_{+(\hat \ell,d)}^{ij}
      &\bar A_{-(\hat \ell+2,d)}^{ij}
  \end{pmatrix}\\
  &\times 
  \begin{pmatrix}
\square-g\hat f \tilde N^{ij} [(1+\xi)\hat \ell+2+\xi]
      &
      g\hat f \tilde N^{ij}(1-\xi)\sqrt{(\hat \ell+1)(\hat \ell+2)}\\
      g\hat f \tilde N^{ij}(1-\xi)\sqrt{(\hat \ell+1)(\hat \ell+2)}
      &\square - g\hat f \tilde N^{ij}[(1+\xi)(\hat \ell+2)-1]
  \end{pmatrix}
        \begin{pmatrix}
            A_{+(\hat \ell,d)}^{ij}\\ A_{-(\hat \ell+2,d)}^{ij}
        \end{pmatrix},
        \notag
\end{align}
where the sum over $d$ is implied.  Taking suitable values of the
gauge parameter $\xi$ and $\tilde N^{ij}$ for the $SU(2)$ case, the
above expression is consistent with similar equations shown
in~\cite{Buchmuller}.  Diagonalizing the Lagrangian in
eq.~\eqref{lag4Dam3} by using the following orthogonal matrix,
\begin{equation}
 \frac{1}{\sqrt{2\hat \ell+3}}
  \begin{pmatrix}
      \sqrt{\hat \ell+2}
      &
      \sqrt{\hat \ell+1}\\
      -\sqrt{\hat \ell+1}
      &\sqrt{\hat \ell+2}    
  \end{pmatrix},
\end{equation}
we finally obtain the masses for $A_{\pm(\hat \ell,d)}^{ij}$: 
\begin{align}
  M_{\rm ph}^2(A_{-(0,d)}^{ij})
  &=  2g\hat f \tilde N^{ij}(-1/2),
\\
  M_\xi^2(A_{+(0,d)}^{ij})
  &=  2g\hat f \tilde N^{ij}\xi(1/2),
\\
  M_{\rm ph}^2(A_{\pm(\hat \ell,d)}^{ij})
  &=  2g\hat f \tilde N^{ij}(\hat \ell+1/2),
  \qquad \hat \ell\geq 1,\\
  M_\xi^2(A_{\pm(\hat \ell,d)}^{ij})
  &=  2g\hat f \tilde N^{ij}\xi(\hat \ell+1/2),
  \qquad \hat \ell\geq 1, 
\end{align}
for $\tilde N^{ij}>0$. We also have the same mass spectrum for the
$\tilde N^{ij}<0$ case.  The scalars $A_{-(0,d)}^{ij}$ are tachyonic,
and all of the other physical scalars are massive. Half of the masses
depend on the gauge fixing parameter, corresponding to the masses for
would-be Goldstone modes. Their degrees of freedom will be absorbed by
the infinite tower of vector fields, leading them to become massive.

\subsection{Vanishing one-loop potentials for NG bosons}
\label{subsec:potNG}
As discussed in section~\ref{sec:massAmk}, NG bosons appear related to
the breaking of the translational symmetry with non-vanishing
flux. The translational invariance is realized non-linearly, under
which NG bosons shift~\cite{Buchmuller,B1}. In our setup, the NG
bosons are identified to some combinations of the zero modes
$A_{m(0,0)}^k$.  More precisely, for a given set of $N^k$ in
eq.~\eqref{qcondhatb1}, the Wilson line phase degrees of freedom along
the Cartan generator $N^k\hat H_k$ are the NG bosons.

From the tree-level mass spectrum, we can understand that one-loop
corrections do not induce masses and potentials for the NG bosons, as
follows. Remind that we have treated the Wilson line phases $\bm v_m$
in eq.~\eqref{bmnewg2} as some parts of the boundary condition. On the
other hand, as explained in section~\ref{sec:bcT2}, they can also be
treated as the VEVs as in eq.~\eqref{bgconst1}. Both treatments are
gauge equivalent and yield the same mass spectrum in the 4D effective
theory. In general, VEVs of the Wilson line phases are not determined
at tree level but may be fixed by effective potentials generated by
quantum corrections. To obtain the effective potential for the Wilson
line phases, which are now considered to be dynamical variables, we
can formally evaluate the path integral using the loop expansion,
although the tachyonic states exist.  The obtained effective
potentials for the phases are mainly determined by the dependence of
the Wilson line phases on the mass spectrum in the 4D effective
theory. The crucial point is that the Wilson line phases corresponding
to the NG bosons, along the generator $N^k\hat H_k$, completely
disappear from the mass spectrum. This follows from the fact that any
fields coupled to the Wilson line phases corresponding to the NG
bosons are also coupled to the flux and have masses independent of the
Wilson line phases. Thus, Wilson line phases along $N^k\hat H_k$ have
completely flat potential at one-loop level as expected. The other
Wilson line phases have generally non-vanishing one-loop potentials
and become massive.  However, the result may be disturbed by tachyonic
states, which are discussed in the next section.

\section{Phenomenological implications}
\label{sec:pheno}
Non-vanishing flux backgrounds in $SU(n)$ gauge theory on
${\cal M}^4\times T^2$ can give a variety of mass spectra in
low-energy effective theories, as shown above. In this section, we
discuss the phenomenological implications of the magnetized torus. A
crucial feature of this setup, as shown in section~\ref{sec:ammasswt},
is the existence of tachyonic
states~\cite{Bachas,Cremades,Buchmuller}. The tachyonic states appear
independently of the gauge parameter $\xi$ and imply that the
background configuration is unstable; thus, they are expected to
evolve non-zero VEVs. The tachyon condensation may induce further
breaking or restoration of gauge symmetries. Otherwise, the tachyonic
states have to be stabilized by some mechanism.

First, we focus on stabilizing the tachyons by considering additional
contributions to the masses of these tachyonic states with the help of
Wilson line phases. From a theoretical viewpoint, extending the
compactified dimensions by more than two is an interesting
possibility. In flux compactifications of superstring theories, where
the extra dimensions are usually more than two, stabilization of
tachyons by Wilson line phases is often
adopted~\cite{Buchmuller:2019zhz}. For example, let us assume a flat
4D torus as the extra dimension. As we have shown, flux backgrounds
generated by $\bm B_5$ and $\bm B_6$ break the gauge symmetry $G$ to
$H$. In addition, we can consider non-trivial configurations of Wilson
line phases included in the backgrounds of the other extra-dimensional
components, denoted by $\bm B_7$ and $\bm B_8$. If the flux background
and the Wilson line phases are in the same direction in the
representation space of $G$, all of the tachyonic states appearing in
the Landau level excitations receive additional contributions to their
masses from the Wilson line phases. Then, the tachyonic states can be
eliminated from the tree-level mass spectrum.

Let us now briefly discuss a possible mass spectrum in an
eight-dimensional setup. As in the previous sections, the tree-level
mass spectrum is determined by the eigenvalues of the differential
operator $-(D_{m'}^{(ij)})^2$, with $m'=5,6,7,8$. The tree-level mass
squared for physical excitations of $A_{\mu}^{ij}$ and
$(A_7^{ij},A_8^{ij})$ are denoted by $M_+^2$, and equivalently
$(A_5^{ij},A_6^{ij})$ are denoted by $M_-^2$. Under a simple setup, if
these fields couple to the flux background in $\bm B_5$ and $\bm B_6$,
the masses can be given by
\begin{align}
  M_\pm^2&=2g\hat f|\tilde N|(\hat \ell\pm 1/2)+
  \left({2\pi}\rho\right)^2\left[(\hat n-\tilde a_7)^2+(\hat m-\tilde a_8)^2\right],
\end{align}
where $\hat f$ is a flux unit, $\tilde N$ is an integer, and $\rho$
parametrizes the relative size of the compact extra dimensions.  The
Wilson line phases are parametrized by the real degrees of freedom
$\tilde a_7$ and $\tilde a_8$. The non-negative integer $\hat \ell$
labels the Landau level excitations, whereas the integers $\hat n$ and
$\hat m$ label KK modes. In the case with vanishing Wilson line phases
$\tilde a_7=\tilde a_8=0$ (mod 1), negative mass squared appears in
$M_-^2$ for $\hat \ell=\hat n=\hat m=0$. Therefore, non-vanishing
$\tilde a_7$ and $\tilde a_8$ can make the mass squared positive,
stabilizing the tachyons.

Note that the Wilson line phases are regarded as continuous moduli
that parametrize the flat directions of the tree-level potential for
the gauge fields, and hence non-vanishing values for the Wilson line
phases must be set by hand at tree level. This implies that
stabilizing the non-vanishing Wilson line phases arises as an
additional issue of this setup. In general, both the flat and non-flat
directions of the potential for extra-dimensional gauge fields are
disturbed by quantum corrections,\footnote{The quantum corrections on
  non-flat directions in 6D orbifold models are discussed, for
  instance, in~\cite{Scrucca:2003ut}. Also, in supersymmetric
  five-dimensional models, extra-dimensional gauge fields and scalars
  belonging to the vector multiplets have tree-level potentials, whose
  non-flat directions receive quantum corrections~\cite{nonflat5D}.}
and finite potentials for Wilson line phases appear at one-loop level
as discussed in the introduction. In a low-energy limit of an
intersecting D-brane model and its T-dual, the one-loop effective
potentials for Wilson line phases and the vacuum configuration were
examined to discuss the stability of tachyons related to flux
~\cite{Buchmuller:2019zhz}. The system is shown to be driven to a
supersymmetric vacuum, where tachyonic states appear. As a future
study, we expect to derive a vacuum configuration that stabilizes the
tachyons at one-loop level in a more general field-theoretical setup.

Next, we discuss the possibility of tachyon condensation.  If the
tachyons are not stabilized, they are expected to evolve non-vanishing
VEVs.  Justifying the existence of VEVs through tachyonic state
condensation is a non-trivial matter. The effective 4D theory contains
infinite scalar modes, and the full scalar potential is quite
complicated. Since the other massive and massless scalar fields
generally receive a backreaction of VEVs of the tachyonic states, the
scalar potential in the effective 4D theory is hard to be examined
analytically, even at tree level. Still, it could be studied
numerically with approximations~\cite{Alfaro:2006is}.  From another
point of view, their VEVs must also be regarded as a background
configuration of a 6D field since the tachyonic states arise from the
extra-dimensional gauge fields. Thus, besides the potential analysis
in an effective 4D theory, another approach for studying the tachyon
condensation is to examine background configurations of the gauge
field as a 6D field. This approach has revealed the restoration of
supersymmetry, which was previously broken by flux backgrounds in an
intersecting D-brane model~\cite{Buchmuller:2020nnl}. In this work, in
which we discussed a Yang-Mills theory without supersymmetry, the
tree-level EOM seems to prevent non-trivial background configurations
that correspond to the tachyon condensation. Some modifications of the
EOM through quantum corrections or extensions of the model are
expected to provide this condensation. While a more comprehensive
analysis of tachyon condensation is necessary, it is still worthwhile
to discuss the expected consequences when assuming the condensation.

\begin{table}
    \centering
    \begin{tabular}{ccc}
      \hline\hline
      \multicolumn{3}{c}{$k=1,2$ and $(i,j)=(1,2)$; \qquad representation of $SU(2)_{U(1)}$: $\bm 3_0\oplus
      \bm 1_0$} \\
      4D fields & masses &
      \\\hline
      $A_{\mu(0,0)}^k,A_{\mu(0,0)}^{ij}$ 
      &0& massless vectors\\
      $A_{m(0,0)}^k,A_{m(0,0)}^{ij}$ 
      &$0$ & massless scalars\\
      $A_{\mu(\hat n,\hat m)}^k,A_{\mu(\hat n,\hat m)}^{ij}$ 
      &$(2\pi)^2\left[\hat n^2+
        \tau_{\rm I}^{-2}(\hat m
        -\hat n\tau_{\rm R})^2\right]$ & massive vectors\\
      $A_{m(\hat n,\hat m)}^k,A_{m(\hat n,\hat m)}^{ij}$ 
      &$(2\pi)^2\left[\hat n^2+
        \tau_{\rm I}^{-2}(\hat m
        -\hat n\tau_{\rm R})^2\right]$ & massive scalars\\
      $A_{m(\hat n,\hat m)}^k,A_{m(\hat n,\hat m)}^{ij}$ 
      &$\xi(2\pi)^2\left[\hat n^2+
        \tau_{\rm I}^{-2}(\hat m
        -\hat n\tau_{\rm R})^2\right]$ & would-be Goldstone modes
      \\\hline\hline &&\\\hline\hline
      \multicolumn{3}{c}{$(i,j)=(1,3),(2,3)$; \qquad representation of $SU(2)_{U(1)}$: $\bm 2_3\oplus
      \bm 2_{-3}$} \\
      4D fields & masses &
      \\\hline
      $A_{-,0,d}^{ij}$ &$-3g\hat f $ & tachyonic states\\
      $A_{\mu,\hat \ell,d}^{ij}$ 
      &$6g\hat f (\hat \ell+1/2)$ & massive vectors\\
      $A_{\pm,\hat \ell,d}^{ij}$ &$6g\hat f (\hat \ell+3/2)$ & massive scalars\\
      $A_{+,0,d}^{ij},A_{\pm,\hat \ell,d}^{ij}$ &$6g\hat f \xi(\hat \ell+1/2)$ & would-be Goldstone modes\\
      \hline\hline
    \end{tabular}
    \caption{Tree level mass spectrum in the $SU(3)$
      model with the flux $(N^1,N^2)=(1,2)$ and vanishing Wilson line phases.
      In the table, $\hat n,\hat m,\hat \ell\in \mathbb Z_{\geq 0}$ and 
      $(\hat n,\hat m)\neq 0$ are implied. The subscript $d$ takes 0 to 2. We denote the representations of the 4D fields under
      $H$ as $SU(2)_{U(1)}$.  }
\label{tablesu3}
\end{table}

As an illustrative toy model, we discuss a 6D $SU(3)$ theory with the
flux configuration $(N^1,N^2)=(1,2)$, which gives $\tilde N^{12}=0$
and $\tilde N^{13}=\tilde N^{23}=3$. This flux breaks $G=SU(3)$ to its
subgroup $H=SU(2)\times U(1)$. Although the non-vanishing continuous
Wilson line phases break $SU(2)$ into $U(1)$, let us consider the case
with vanishing Wilson line phases, $\tilde v^{ij}=0$, for
simplicity. The mass spectrum in this setup is shown in
Table~\ref{tablesu3}. There appear massless 4D gauge fields that
transform under $H$ as the adjoint, $\bm 3_0\oplus \bm 1_0$, with the
representation denoted as $SU(2)_{U(1)}$. There are also massless
adjoint scalars and tachyonic states that belong to the generator of
$G/H$, i.e., the $SU(2)$ doublets $\bm 2_3\oplus \bm 2_{-3}$. The
remaining fields are either massive or would-be Goldstone modes. If
the tachyonic states develop constant VEVs, keeping the other part of
the backgrounds invariant, unitary transformations make the VEVs
parametrized by a real parameter $\phi$ as
$\vev{\bm 2_3}^T=\vev{\bm 2_{-3}}=(0,\phi)$, where $T$ represents the
transpose. For a non-zero value of $\phi$, some of the massless
adjoint fields acquire masses, resulting in the symmetry breaking
$SU(2)\times U(1)\to U(1)'$. This presents the possibility of the EWSB
triggered by the Nielsen-Olsen type instability~\cite{Nielsen:1978rm},
where the SM Higgs scalar originates from the lowest Landau level of
the extra-dimensional gauge fields. However, various aspects such as
vacuum stability, the matter sector, and other details of the model
must be examined in detail.

With enlarged gauge symmetries, flux backgrounds combined with tachyon
condensation may provide several possibilities for symmetry-breaking
patterns.  For example, in the $SU(5)$ case with
$(N^1,N^2,N^3,N^4)=(2,4,6,3)$, the flux breaks $G=SU(5)$ to its
subgroup $H=SU(3)\times SU(2)\times U(1)$. For vanishing Wilson line
phases, there appear tachyonic states belonging to the generator of
$G/H$, namely $(\bm 3,\bm 2)_5$, where the representation is written
as $(SU(3),SU(2))_{U(1)}$. Their VEVs can induce the symmetry breaking
$SU(3)\times SU(2)\times U(1)\to SU(2)_{\rm D}\times U(1)'$, where
$SU(2)_{\rm D}$ is the diagonal part of
$SU(2)'\times SU(2)\subset SU(3)\times SU(2)$, and $U(1)'$ is the
linear combination of a $U(1)$ generator and a diagonal generator in
$SU(3)/SU(2)'$. The resultant symmetry may be identified as the
electroweak symmetry. Flux backgrounds and tachyon condensation can
also be utilized to break unified gauge symmetries in GUT models. If
we consider GUT models with $G=SU(7)$, $SU(8)$, $SO(10)$, or $E_6$,
flux backgrounds can break $G$ to $H=SU(5)\times G'$, and the
condensation of tachyons belonging to $G/H$ seems to provide similar
symmetry-breaking patterns in known five-dimensional
models~\cite{Kojima:2017qbt}. In this case, the Wilson line phases
belonging to $G/H$ induce the symmetry breaking into the SM gauge
symmetry, in contrast to the tachyon condensation.

Finally, we briefly mention matter fields under flux backgrounds. It
is well known that when fermion fields are present, the lowest Landau
level excitations of the fermions that couple to flux backgrounds
become massless states and give rise to chiral mass spectra. The
strength of the coupling is expressed by the integer $\tilde N^{ij}$
in eq.~\eqref{deftilN} and determines the generation number of the
chiral fermions at a low-energy regime. While we have mainly focused
on $SU(n)$ so far, examining the example of $E_6$ GUT models can
provide insight into the generation structure. If there is a flux
background along the $U(1)_X$ generator, which is a part of the
maximal subgroup $SO(10)\times U(1)_X$ of $E_6$, then the flux can
break $E_6$ into $SO(10)\times U(1)_X$. Furthermore, this symmetry can
be broken down to $SU(5)\times U(1)'$ or the SM gauge group with the
help of tachyons or VEVs of Wilson line phases. In terms of matter
fields, we can incorporate a 27--plet of $E_6$, which is decomposed
into the representations of $SO(10)_{U(1)_X}$ as
$\bm{27}\to \bm{16}_{-1}+\bm{10}_2+\bm{1}_{-4}$. When the generation
of light modes from $\bm{16}_{-1}$ is three, which have fermion
content suitable to the SM matter fields and right-handed neutrinos,
the generations from $\bm{10}_2$ and $\bm{1}_{-4}$ are six and twelve,
respectively. We note that fermions in $\bm{10}_2$ and $\bm{1}_{-4}$
are vector-like under both $SU(5)\times U(1)'$ and the SM gauge group,
and thus are expected to acquire masses during the symmetry breaking
to the SM. Light chiral fermions could be mixed states from three
generations of light fields from $\bm{16}_{-1}$ and six generations of
light fields from $\bm{10}_2$. Additionally, the matter contents may
be regarded as an extension of the twisted flavor structure discussed
in $E_6$ GUT
models~\cite{E6twist,Bando:2000gs,Inoue:2007ed}. Comprehensive studies
of GUT models, including the prediction of the flavor structure, are
interesting issues left for future exploration.

\section{Conclusions}
\label{sec:conc}
In this study, we have investigated an $SU(n)$ gauge theory in a 6D
spacetime with a constant magnetic flux in the two-dimensional torus.
By analyzing the classical equations of motion, a general form of
consistent background configurations incorporating both the magnetic
flux and Wilson line phases has been obtained.  Then, we have derived
the appropriate boundary conditions for fields associated with the
discrete translations on the torus. There are many possible setups
since the boundary conditions adjust to changes in the background
configuration induced by gauge transformations. We have chosen a gauge
that incorporates the Wilson line phases into the twist matrices. In
addition, we have performed a standard $R_\xi$ gauge fixing,
demonstrating the dependence of the masses for the KK modes on the
gauge fixing parameter.

Keeping the gauge parameter and the background configurations
arbitrary, the complete expressions of the tree-level mass spectrum in
the effective low-energy theory is newly obtained for a general
$SU(n)$ case.  Our analysis confirmed the existence of tachyonic
modes, which appear independently of the gauge fixing parameter $\xi$
or the background configuration of Wilson line phases with flux. Some
of the remaining scalar fields coupled to the flux were found to be
massive, while others that showed dependence on $\xi$ were identified
as would-be Goldstone modes. Consequently, the degrees of freedom of
the latter are absorbed by the infinite tower of 4D vector fields,
rendering them massive. As expected, only the masses of flux-blind
scalar fields in the non-Cartan directions receive contributions from
the Wilson line phases. We have also discussed vanishing one-loop
potentials for NG bosons related to the violation of the translational
symmetry. The one-loop potential for the NG bosons is not generated
since the VEVs of Wilson line phases corresponding to the NG bosons do
not appear in the mass parameters for any 4D modes.

Based on our findings, we have discussed the phenomenological
implications associated with stabilization or condensation of the
tachyonic states.  The elimination of tachyonic modes through Wilson
line phases is still viable by increasing the number of extra
dimensions, where the stabilization of the vacuum at a quantum level
provides an interesting topic for future research. Also, we have
discussed implications related to the tachyon condensation.  In this
setup, it is expected to be possible to generate various mass spectra
and explore different patterns of symmetry breaking.  Therefore,
extending the gauge group of the magnetic torus setup to
simply-connected ones enables our results to be applied in the
construction of diverse models within GUT and GHU.  The exploration of
phenomenologically viable models, considering the prediction of the
flavor structure through the inclusion of matter fields, is another
intriguing topic left for future studies.

\bigskip \bigskip

\begin{center}
{\bf Acknowledgement}
\end{center}
The authors would like to thank T.~Hirose and N.~Maru for helpful
communication and also thank N.~Yamatsu and H.~Otsuka for the fruitful discussion
and comments. The work of Y.O. is supported in part by the Kyushu
University Leading Human Resources Development Fellowship Program.
The work of C.T. is supported by the Ministry of Education, Culture,
Sports, Science and Technology (MEXT) of Japan.
\bigskip \bigskip
\appendix
%

\section{Surface terms}
\label{app:surf}
We discuss surface terms related to integration by parts in our
setup. We examine the background covariant derivative in
eq.~\eqref{bgdfdef1}, which depends on the torus coordinates and is
denoted by ${\cal D}_m(z)$.  With the background configurations and
the twist matrices in eqs.~\eqref{bmnewg1} and~\eqref{bmnewg2}, one
obtains the boundary conditions for ${\cal D}_m(z)$ as
\begin{align}\label{covbc1}
  {\cal D}_m({\cal T}_pz)=T_p(z){\cal D}_m(z)T_p^\dag(z).
\end{align}

Let $\bm \phi(z), \bm \phi'(z)\in su(n)$ be general fields in the
adjoint representation that obey the following relations:
\begin{align}\label{bcphiex1}
  {\cal D}_m\bm \phi=(\der_m-ig\ad{\bm B_m})\bm \phi, \qquad
  \bm \phi({\cal T}_pz)=T_p(z)\bm \phi(z)T_p^\dag (z).
\end{align}
For $\bm \phi'(z)$, the same relations are applied. In this case, one sees that 
\begin{align} 
  \Tr[({\cal D}_m\bm \phi)({\cal D}_m\bm \phi')]=
  -\Tr[\bm \phi({\cal D}_m{\cal D}_n\bm \phi')]+\der_m\Tr[\bm \phi({\cal D}_n\bm \phi')],
\end{align}
where the last term gives the vanishing contribution in the action as
\begin{align}
  \int_{{\cal M}^4}d^4x\int_{T^2}d^2x~\der_m\Tr[\bm \phi({\cal D}_n\bm \phi')]
  \propto \big[\Tr[\bm \phi({\cal D}_n\bm \phi')]\big]_z^{{\cal T}_pz}=0.
\end{align}
Here, we have used the boundary conditions in eqs.~\eqref{covbc1}
and~\eqref{bcphiex1} and the cyclic property of traces.

Using similar discussions as the above, we obtain
eqs.~\eqref{L1201}--\eqref{L1203},~\eqref{gflag3},~\eqref{qlagAmu1},
and~\eqref{qlagAm1} without contributions from surface terms.

\section{Mode Functions}
\label{appendix:modefunction}
In section~\ref{sec:Aijlandaumass}, we showed the explicit form of the
mode functions without derivation.  In this appendix, we show the
derivation.

\subsection{Zero Mode Functions}
First, let us derive the zero-mode function $\zeta_{0,d}$ appearing in
eq.~\eqref{modefunction}.  For the case with $N^{ij}>0$, we have to
solve the differential equation
\begin{align}
    \label{eqn:difeqpN}
    \left(\bar{\partial}_z+\frac{g\hat{f}\tilde{N}}{4}(z-\gamma \bar{z}) \right)\zeta_{0,d}(z)=0,
\end{align}
with the boundary conditions in eqs.~\eqref{bczetaijT5} and
\eqref{bczetaijT6}.  Here, we omit the upper script $ij$ for
simplicity of notation.  A solution of eq.~\eqref{eqn:difeqpN} is
given by
\begin{align}
    \zeta_{0,d}(z)
    =
    Ce^{-\frac{g\hat{f}\tilde{N}}{4}(z\bar{z}-\frac12\gamma \bar{z}^2)}
    \tilde{\zeta}_{0,d}(z),
\end{align}
where $C$ is a constant, and the function $\tilde{\zeta}_{0,d}(z)$
depends only on $z$ and does not depend on $\bar{z}$.  To determine
the function $\tilde{\zeta}_{0,d}(z)$ completely, we have to impose
the boundary conditions.  It is convenient to introduce a new function
$\chi_{0,d}$ by
\begin{align}
    \tilde{\zeta}_{0,d}(z)
    =
    e^{\frac{g\hat{f}\tilde{N}}{4}(1-\frac{\gamma}{2})z^2}
    \chi_{0,d}(z), 
\end{align}
to express the solution as
\begin{align}
    \zeta_{0,d}(z)
    &=
    Ce^{-\frac{g\hat{f}\tilde{N}}{4}(z\bar{z}-z^2+\frac{\gamma}{2}(z^2- \bar{z}^2))}
    \chi_{0,d}(z).
\end{align}

From the conditions in eqs.~\eqref{bczetaijT5} and \eqref{bczetaijT6},
we obtain the boundary conditions for $\chi_{0,d}(z)$ as
\begin{align}
    \label{bcchiijT5}
    \chi_{0,d}(z+1)&=e^{-2\pi i\tilde{a}_5}\chi_{0,d}(z),
    \\
    \label{bcchiijT6}
    \chi_{0,d}(z+\tau)
    &=
    e^{-2\pi i \left(\tau_{\mathrm{R}} \tilde{a}_5+\tau_{\mathrm{I}} \tilde{a}_6\right)}
    e^{-\pi i\tilde{N}\tau}
     e^{\gamma \pi i \tilde{N} \tau_{\rm R}}
    e^{-2\pi i\tilde{N}z}
    \chi_{0,d}(z), 
\end{align}
where we have used
$\hat{f}=2\pi/(g\mathcal{V}_{T^2})=2\pi/(g\tau_{\rm I})$ to obtain the
second relation.  The function $\chi_{0,d}$ satisfying these boundary
conditions can be expressed by the Jacobi theta
function~\cite{Cremades}.  The Jacobi theta function
$\Jthe{a}{b}{z'}{\tau'}$ is defined by
\begin{align}
  \Jthe{a}{b}{z'}{\tau'}=\sum_{n\in \mathbb Z} e^{i\pi\tau'(n+a)^2}e^{2\pi i(n+a)(z'+b)}, \qquad
  a,b\in\mathbb R, \qquad z',\tau' \in  \mathbb C, 
\end{align}
which satisfies 
\begin{align}\label{eqn:JTbc01}
    \Jthe{a}{b}{z'+k}{\tau'}&=e^{2\pi ika}\Jthe{a}{b}{z'}{\tau'}, \qquad k\in \mathbb Z, \\
  \label{eqn:JTbc02}
  \Jthe{a}{b}{z'+ \tau'}{\tau'}&=
                               e^{-i\pi\tau'} e^{- 2\pi i( z'+b)}\Jthe{a}{b}{z'}{\tau'}.
\end{align}
Using it, we can express $\chi_{0,d}$ as 
\begin{align}
    \label{eqn:chivartheta}
    \chi_{0,d}(z)
    =
    \vartheta
    \left[
    \begin{array}{c}
    -(\tilde{a}_5+d)/\tilde{N}
      \\
      \tau_{\rm R}\tilde a_5+\tau_{\rm I}\tilde a_6-\gamma\tau_{\rm R}\tilde{N}/2
    \end{array}
    \right]
    (\tilde{N}z|\tilde{N}\tau).
\end{align}
From eqs.~\eqref{eqn:JTbc01} and~\eqref{eqn:JTbc02} , we can show that
eq.~\eqref{eqn:chivartheta} satisfies the boundary conditions in
eqs.~\eqref{bcchiijT5} and~\eqref{bcchiijT6}.

Finally, the zero mode function is expressed as
\begin{align}\label{eqn:zeromode}
    \zeta_{0,d}(z)
    =
    C
    e^{-\frac{\pi \tilde{N}}{2\tau_{\rm I}}(z\bar{z}-zz+\frac{\gamma}{2}(zz- \bar{z}\bar{z}))}
    \vartheta
    \left[
    \begin{array}{c}
    -(\tilde{a}_5+d)/\tilde{N}
      \\
      \tau_{\rm R}\tilde a_5+\tau_{\rm I}\tilde a_6-\gamma\tau_{\rm R}\tilde{N}/2
    \end{array}
    \right]
    (\tilde{N}z|\tilde{N}\tau).
\end{align}
The constant $C$ is determined by a normalization condition.  Here, we
impose
\begin{align}
    \int_{\mathcal{V}_{T^2}}dx^5dx^6\ 
    \bar{\zeta}_{0,d}(z)
    \zeta_{0,d'}(z)
    =
    \delta_{d,d'},
\end{align}
which yields
\begin{align}
    C=\left(\frac{2\pi \tilde N}{\tau_{\rm I}} \right)^{1/4}.
\end{align}

\subsection{Excited Mode functions}
The excited mode functions are obtained by
\begin{align}
    \zeta_{\hat \ell,d}(z)
    =
    \frac{1}{\sqrt{\hat \ell !}}
    (\hat{a}^\dagger)^{\hat \ell}
    \zeta_{0,d}(z),
\end{align}
where $\hat{a}^\dagger$ is given by
\begin{align}
    \hat{a}^\dagger
    =
    \Pi_z
    =
    -i\sqrt{\frac{\tau_{\rm I}}{2\pi \tilde{N}}}
    \left[
    \partial_z-\frac{\pi \tilde{N}}{2\tau_{\rm I}}(\bar{z}-\gamma z)
    \right],
\end{align}
for $\tilde{N}^{ij}>0$.
Let us show that the excited mode functions are given by
\begin{align}
    \label{eqn:gammaexmodes}
\begin{aligned}
    \zeta_{\hat \ell,d}(z)
    =
    \frac{C}{2^{\hat \ell}\sqrt{\hat \ell !}}
    \sum_{n=-\infty}^{+\infty}
    &Z_{\hat \ell}(w_{n,d}(z))
    e^{-\frac{\pi \tilde{N}}{2\tau_{\rm I}}(z\bar{z}-zz+\frac{\gamma}{2}(zz- \bar{z}\bar{z}))}
    \\
    &
    \times
    e^{i\pi \tilde{N}\tau(n-(\tilde a_5+d)/{\tilde{N}})^2}
    e^{2\pi i\tilde{N}(n-(\tilde a_5+d)/{\tilde{N}} )
      (z+(\tau_{\rm R}\tilde a_5+\tau_{\rm I}\tilde a_6)/\tilde N-\gamma\tau_{\rm R}/2)}. 
\end{aligned}
\end{align}
Here, the function $w_{n,d}(z)$ is defined by 
\begin{align}
    w_{n,d}(z)
    =
    \sqrt{\frac{2\pi \tilde N}{\tau_{\rm I}}}
    \left[
    \frac{z-\bar{z}}{2i}
    +\tau_{\rm I}\left(n-\frac{\tilde a_5+d}{\tilde{N}} \right)
    \right], 
\end{align}
and $Z_{\hat \ell}$ is defined by
\begin{gather}
    Z_{0}(w_{n,d})=1,\qquad 
    Z_{\hat \ell+1}(w_{n,d})
    =
    2w_{n}Z_{\hat \ell}(w_{n,d})
    -
    \frac{d Z_{\hat \ell}(w_{n,d})}{d w_{n,d}}
    \label{eqn:diffeqforzell}.
\end{gather}

To derive the excited mode functions, let us express the zero-mode
function in eq.~\eqref{eqn:zeromode} as
\begin{gather}
    \zeta_{0,d}(z)
    =
    C\sum_{n=-\infty}^{+\infty}Z_{0}(w_{n,d})F_{n,d}(z),
    \label{eqn:zerozfn}
    \\
    F_{n,d}(z)
    \equiv
    e^{-\frac{\pi \tilde{N}}{2\tau_{\rm I}}(z\bar{z}-zz+\frac{\gamma}{2}(zz- \bar{z}\bar{z}))}
    e^{i\pi \tilde{N}\tau(n-(\tilde a_5+d)/{\tilde{N}})^2}
    e^{2\pi i(n-(\tilde a_5+d)/{\tilde{N}} )    (\tilde{N}z+
\tau_{\rm R}\tilde a_5+\tau_{\rm I}\tilde a_6-\gamma\tau_{\rm R}\tilde{N}/2)}.
\end{gather}
By using the expression above and 
\begin{align}
    \zeta_{1,d}(z)
    =
    \frac{1}{\sqrt{1!}}\hat a^\dagger\zeta_{0,d}(z)
    =
    -i\sqrt{\frac{\tau_{\rm I}}{2\pi \tilde{N}}}
    \left[
    \partial_z-\frac{\pi \tilde{N}}{2\tau_{\rm I}}(\bar{z}-\gamma z)
    \right]
    \zeta_{0,d}(z), 
\end{align}
the first-excited mode is written as 
\begin{align}
    \label{eqn:1stexfromzero}
    \zeta_{1,d}(z)
    =
    -iC\sqrt{\frac{\tau_{\rm I}}{2\pi \tilde{N}}}
    \sum_n
    \left[
    (\partial_zZ_{0})F_{n,d}
    +
    Z_{0}(\partial_zF_{n,d})
    -\frac{\pi \tilde{N}}{2\tau_{\rm I}}(\bar{z}-\gamma z)
    Z_{0}F_{n,d}
    \right].
\end{align}
The last two terms are rearranged as 
\begin{align}
    &Z_{0}(\partial_zF_{n,d})
    -\frac{\pi \tilde{N}}{2\tau_{\rm I}}(\bar{z}-\gamma z)
    Z_{0}F_{n,d}
    \nonumber
    \\
    &=
    \frac{2\pi \tilde{N}}{\tau_{\rm I}}i
    \left(\frac{z-\bar{z}}{2i}+\tau_{\rm I} \left(n-\frac{g\tilde{v}_5}{2\pi \tilde{N}}-\frac{d}{\tilde{N}} \right) \right)
    Z_{0}F_{n,d}
      =
    i\sqrt{\frac{2\pi \tilde{N}}{\tau_{\rm I}}}w_{n,d}Z_{0}F_{n,d},
\end{align}
and the first term can be calculated as
\begin{align}
    (\partial_z Z_{0})F_{n,d}
    =
    \frac{\partial w_{n,d}}{\partial z}
    \frac{d Z_{0}}{d w_{n,d}}
    F_{n,d}
    =
    \frac{1}{2i}
    \sqrt{\frac{2\pi \tilde{N}}{\tau_{\rm I}}}
    \frac{d Z_{0}}{d w_{n,d}}
    F_{n,d}.
\end{align}
By using these results, eq.~\eqref{eqn:1stexfromzero} becomes
\begin{align}
    \zeta_{1,d}(z)
    =
    \frac{C}{2}
    \sum_{n=-\infty}^{+\infty}
    \left(2w_{n,d}(z)Z_{0}(w_{n,d})-\frac{d Z_0(w_{n,d})}{d w_{n,d}} \right)F_{n,d}(z).
\end{align}
Identifying the square bracket with $Z_1$, i.e.,
\begin{align}
    Z_1=2w_{n,d}Z_{0}-\frac{d Z_0}{d w_{n,d}},
\end{align}
the first excited mode can be written as
\begin{align}
    \zeta_{1,d}(z)
    =
    \frac{C}{2}
    \sum_{n=-\infty}^{+\infty}
    Z_1(w_{n,d}(z))F_{n,d}(z).
\end{align}
Therefore, eq.~\eqref{eqn:gammaexmodes} is correct for $\hat \ell=1$.

Let us assume eq.~\eqref{eqn:gammaexmodes} for $\hat \ell=k$ and show it for $\hat \ell=k+1$.
Using  
\begin{align}
    \zeta_{k+1,d}(z)
    =
    \frac{1}{\sqrt{k+1}}a^\dagger \zeta_{k,d}(z),
\end{align}
and repeating the same calculation we did for the $\hat \ell=0$ case,
we get
\begin{align}
    \zeta_{k+1,d}(z)
    =
    \frac{C}{2^{k+1}\sqrt{(k+1)!}}
    \sum_{n=-\infty}^{+\infty}
    \left(2w_{n,d}Z_k-\frac{dZ_k}{dw_{n,d}} \right)F_{n,d}(z).
\end{align}
Hence, eq.~\eqref{eqn:gammaexmodes} is satisfied for $\hat \ell=k+1$.
Therefore, the assumption holds for any $\hat \ell$.

The function $Z_{\hat \ell}(w_n)$, which satisfies the differential
equation in eq.~\eqref{eqn:diffeqforzell}, can be expressed by Hermite
polynomials $H_{\hat \ell}(x)$.  Therefore, we obtain the expression
in eq.~\eqref{eqn:gammaexmodes} for $\tilde{N}^{ij}>0$.  In the same
way, we can obtain the explicit form of the mode function for
$\tilde{N}^{ij}<0$ as
\begin{align}
\begin{aligned}
    \zeta_{\hat \ell,d}(z)
    =
    \frac{C}{2^{\hat \ell}\sqrt{\hat \ell !}}
    \sum_{n=-\infty}^{+\infty}
    &H_{\hat \ell}(w_{n,d}(z))
    e^{-\frac{\pi \tilde{N}}{2\tau_{\rm I}}(\bar{z}z-\bar{z}\bar{z}+\frac{\gamma}{2}(\bar{z}\bar{z}- zz))}
    \\
    &
    \times
    e^{-i\pi \tilde{N}\bar{\tau}(n-(\tilde a_5+d)/{\tilde{N}})^2}
    e^{-i2\pi\tilde{N}(n-(\tilde a_5+d)/{\tilde{N}} )(\bar{z}+(\tau_{\rm R}\tilde a_5+\tau_{\rm I}\tilde a_6)/\tilde N-\gamma\tau_{\rm R}/2)}.
\end{aligned}
\end{align}
These functions correctly satisfy the boundary conditions in
eqs.~\eqref{bczetaijT5} and~\eqref{bczetaijT6} and the orthogonal
relation in eq.~\eqref{orthogonal}.

\bigskip\bigskip


\begin{thebibliography}{99}


\bibitem{GG}
    H.~Georgi and S.~L.~Glashow,
    ``Unity of All Elementary Particle Forces,''
    Phys.\ Rev.\ Lett.\  {\bf 32}, (1974), 438.
\bibitem{manton}
    N.~S.~Manton,
    ``A New Six-Dimensional Approach To The Weinberg-Salam Model,''
    Nucl.\ Phys.\  B {\bf 158}, (1979), 141.
\bibitem{fair}
    D.~B.~Fairlie,
    ``Higgs' Fields And The Determination Of The Weinberg Angle,''
    Phys.\ Lett.\  B {\bf 82}, (1979), 97.
\bibitem{hosotani1}
  Y.~Hosotani,
  ``Dynamical Mass Generation by Compact Extra Dimensions,''
  Phys.\ Lett.\ B {\bf 126}, (1983), 309.
\bibitem{Hosotani:1988bm}
    Y.~Hosotani,
    ``Dynamics of Nonintegrable Phases and Gauge Symmetry Breaking,''
    Annals Phys. \textbf{190} (1989), 233.
\bibitem{Hatanaka:1998yp}
    H.~Hatanaka, T.~Inami and C.~S.~Lim,
    ``The Gauge hierarchy problem and higher dimensional gauge theories,''
    Mod. Phys. Lett. A \textbf{13} (1998), 2601-2612
    [arXiv:hep-th/9805067].
\bibitem{Hall:2001zb}
    L.~J.~Hall, Y.~Nomura and D.~Tucker-Smith,
    ``Gauge Higgs unification in higher dimensions,''
    Nucl. Phys. B \textbf{639} (2002), 307-330
    [arXiv:hep-ph/0107331].
\bibitem{Antoniadis:2001cv}
    I.~Antoniadis, K.~Benakli and M.~Quiros,
    ``Finite Higgs mass without supersymmetry,''
    New J. Phys. \textbf{3} (2001), 20
    [arXiv:hep-th/0108005].
\bibitem{Kubo:2001zc}
    M.~Kubo, C.~S.~Lim and H.~Yamashita,
    ``The Hosotani mechanism in bulk gauge theories with an orbifold extra space S**1 / Z(2),''
    Mod. Phys. Lett. A \textbf{17} (2002), 2249-2264
    [arXiv:hep-ph/0111327].
\bibitem{Csaki:2002ur}
    C.~Csaki, C.~Grojean and H.~Murayama,
    ``Standard model Higgs from higher dimensional gauge fields,''
    Phys. Rev. D \textbf{67} (2003), 085012
     [arXiv:hep-ph/0210133].
\bibitem{Burdman:2002se}
    G.~Burdman and Y.~Nomura,
    ``Unification of Higgs and Gauge Fields in Five Dimensions,''
    Nucl. Phys. B \textbf{656} (2003), 3-22
    [arXiv:hep-ph/0210257].
\bibitem{Gogoladze:2003bb}
    I.~Gogoladze, Y.~Mimura and S.~Nandi,
     ``Gauge Higgs unification on the left right model,''
    Phys. Lett. B \textbf{560} (2003), 204-213
     [arXiv:hep-ph/0301014].
\bibitem{Scrucca:2003ra}
    C.~A.~Scrucca, M.~Serone and L.~Silvestrini,
    ``Electroweak symmetry breaking and fermion masses from extra dimensions,''
    Nucl. Phys. B \textbf{669} (2003), 128-158
     [arXiv:hep-ph/0304220].
\bibitem{Haba:2004qf}
    N.~Haba, Y.~Hosotani, Y.~Kawamura and T.~Yamashita,
    ``Dynamical symmetry breaking in gauge Higgs unification on orbifold,''
    Phys. Rev. D \textbf{70} (2004), 015010
    [arXiv:hep-ph/0401183].
\bibitem{Haba:2004jd}
    N.~Haba and T.~Yamashita,
     ``Dynamical symmetry breaking in gauge Higgs unification of 5-D N=1 SUSY theory,''
    JHEP \textbf{04} (2004), 016
    [arXiv:hep-ph/0402157].
\bibitem{Hosotani:2004ka}
    Y.~Hosotani, S.~Noda and K.~Takenaga,
    ``Dynamical gauge symmetry breaking and mass generation on the orbifold T**2 / Z(2),''
    Phys. Rev. D \textbf{69} (2004), 125014
    [arXiv:hep-ph/0403106].
\bibitem{Hosotani:2004wv}
    Y.~Hosotani, S.~Noda and K.~Takenaga,
    ``Dynamical gauge-Higgs unification in the electroweak theory,''
    Phys. Lett. B \textbf{607} (2005), 276-285
    [arXiv:hep-ph/0410193].
\bibitem{Haba:2004bh}
    N.~Haba, K.~Takenaga and T.~Yamashita,
    ``Higgs mass in the gauge-Higgs unification,''
    Phys. Lett. B \textbf{615} (2005), 247-256
    [arXiv:hep-ph/0411250].
\bibitem{Lim:2007jv}
    C.~S.~Lim and N.~Maru,
    ``Towards a realistic grand gauge-Higgs unification,''
    Phys. Lett. B \textbf{653} (2007), 320-324
    [arXiv:0706.1397].
\bibitem{Kojima:2011ad}
    K.~Kojima, K.~Takenaga and T.~Yamashita,
    ``Grand Gauge-Higgs Unification,''
    Phys. Rev. D \textbf{84} (2011), 051701
    [arXiv:1103.1234].
\bibitem{Yamashita:2011an}
    T.~Yamashita,
    ``Doublet-Triplet Splitting in an SU(5) Grand Unification,''
    Phys. Rev. D \textbf{84} (2011), 115016
    [arXiv:1106.3229].
\bibitem{Hosotani:2015hoa}
    Y.~Hosotani and N.~Yamatsu,
    ``Gauge\textendash{}Higgs grand unification,''
    PTEP \textbf{2015} (2015), 111B01
    [arXiv:1504.03817].
\bibitem{Yamatsu:2015oit}
    N.~Yamatsu,
    ``Gauge coupling unification in gauge\textendash{}Higgs grand unification,''
    PTEP \textbf{2016} (2016) no.4, 043B02
    [arXiv:1512.05559].
\bibitem{Furui:2016owe}
    A.~Furui, Y.~Hosotani and N.~Yamatsu,
    ``Toward Realistic Gauge-Higgs Grand Unification,''
    PTEP \textbf{2016} (2016) no.9, 093B01
    [arXiv:1606.07222].
\bibitem{Kojima:2016fvv}
    K.~Kojima, K.~Takenaga and T.~Yamashita,
    ``Gauge symmetry breaking patterns in an SU(5) grand gauge-Higgs unification model,''
    Phys. Rev. D \textbf{95} (2017) no.1, 015021
    [arXiv:1608.05496].
\bibitem{Kojima:2017qbt}
    K.~Kojima, K.~Takenaga and T.~Yamashita,
    ``The Standard Model Gauge Symmetry from Higher-Rank Unified Groups in Grand Gauge-Higgs Unification Models,''
    JHEP \textbf{06} (2017), 018
    [arXiv:1704.04840].
\bibitem{Hosotani:2017edv}
    Y.~Hosotani and N.~Yamatsu,
    ``Electroweak symmetry breaking and mass spectra in six-dimensional gauge\textendash{}Higgs grand unification,''
    PTEP \textbf{2018} (2018) no.2, 023B05
    [arXiv:1710.04811].
\bibitem{Maru:2019lit}
    N.~Maru and Y.~Yatagai,
    ``Fermion Mass Hierarchy in Grand Gauge-Higgs Unification,''
    PTEP \textbf{2019} (2019) no.8, 083B03
    [arXiv:1903.08359].
\bibitem{Englert:2019xhz}
    C.~Englert, D.~J.~Miller and D.~D.~Smaranda,
    ``Phenomenology of GUT-inspired gauge-Higgs unification,''
    Phys.\ Lett.\  B {\bf 802}, (2020), 135261
    [arXiv:1911.05527 [hep-ph]].
\bibitem{Angelescu:2021nbp}
    A.~Angelescu, A.~Bally, S.~Blasi and F.~Goertz,
    ``Minimal SU(6) gauge-Higgs grand unification,''
    Phys. Rev. D \textbf{105} (2022) no.3, 035026
    [arXiv:2104.07366].
\bibitem{Nakano:2022lyt}
    H.~Nakano, M.~Sato, O.~Seto and T.~Yamashita,
    ``Dirac gaugino from grand gauge-Higgs unification,''
    PTEP \textbf{2022} (2022) no.3, 033B06
    [arXiv:2201.04428].
\bibitem{Kojima:2023mew}
    K.~Kojima, K.~Takenaga and T.~Yamashita,
    ``Grand Gauge-Higgs Unification on $T^2/{\mathbb Z}_3$ via Diagonal Embedding Method,''
    arXiv:2304.05701.
\bibitem{Maru:2006wa}
    N.~Maru and T.~Yamashita,
    ``Two-loop Calculation of Higgs Mass in Gauge-Higgs Unification: 5D Massless QED Compactified on S**1,''
    Nucl. Phys. B \textbf{754} (2006), 127-145
    [arXiv:hep-ph/0603237].
\bibitem{Hosotani:2007kn}
    Y.~Hosotani, N.~Maru, K.~Takenaga and T.~Yamashita,
    ``Two Loop finiteness of Higgs mass and potential in the gauge-Higgs unification,''
    Prog. Theor. Phys. \textbf{118} (2007), 1053-1068
    [arXiv:0709.2844].
\bibitem{Review} 
    M.~R.~Douglas and S.~Kachru,
    ``Flux compactification,''
    Rev. Mod. Phys. \textbf{79} (2007), 733-796
    [arXiv:hep-th/0610102].
\bibitem{Review2} 
    M.~Grana,
    ``Flux compactifications in string theory: A Comprehensive review,''
    Phys. Rept. \textbf{423} (2006), 91-158
    [arXiv:hep-th/0509003].
\bibitem{Generations} 
    E.~Witten,
    ``Some Properties of O(32) Superstrings,''
    Phys. Lett. B \textbf{149} (1984), 351-356.
\bibitem{Bachas} 
    C.~Bachas,
    ``A Way to break supersymmetry,''
    arXiv:hep-th/9503030.
\bibitem{Abe:2008sx}
    H.~Abe, K.~S.~Choi, T.~Kobayashi and H.~Ohki,
    ``Three generation magnetized orbifold models,''
    Nucl. Phys. B \textbf{814}, (2009), 265-292 
    [arXiv:0812.3534].
    \bibitem{Kobayashi:2010an}
    T.~Kobayashi, R.~Maruyama, M.~Murata, H.~Ohki and M.~Sakai,
    ``Three-generation Models from $E_8$ Magnetized Extra Dimensional Theory,''
    JHEP \textbf{05} (2010), 050
    [arXiv:1002.2828].
\bibitem{Abe:2015yva}
    T.~h.~Abe, Y.~Fujimoto, T.~Kobayashi, T.~Miura, K.~Nishiwaki, M.~Sakamoto and Y.~Tatsuta,
    ``Classification of three-generation models on magnetized orbifolds,''
    Nucl. Phys. B \textbf{894}, (2015), 374-406
    [arXiv:1501.02787].
\bibitem{Abe:2015mua}
    H.~Abe, T.~Kobayashi, H.~Otsuka and Y.~Takano,
    ``Realistic three-generation models from SO(32) heterotic string theory,''
    JHEP \textbf{09} (2015), 056
    [arXiv:1503.06770].
\bibitem{Sakamoto:2020pev}
    M.~Sakamoto, M.~Takeuchi and Y.~Tatsuta,
    ``Zero-mode counting formula and zeros in orbifold compactifications,''
    Phys. Rev. D \textbf{102}, (2020) no.2, 025008 
    [arXiv:2004.05570].
\bibitem{Sakamoto:2020vdy}
    M.~Sakamoto, M.~Takeuchi and Y.~Tatsuta,
    ``Index theorem on $T^2/\mathbb{Z}_N$ orbifolds,''
    Phys. Rev. D \textbf{103}, (2021) no.2, 025009 
    [arXiv:2010.14214].
\bibitem{Kobayashi:2022tti}
    T.~Kobayashi, H.~Otsuka, M.~Sakamoto, M.~Takeuchi, Y.~Tatsuta and H.~Uchida,
    ``Index theorem on magnetized blow-up manifold of T2/ZN,''
    Phys. Rev. D \textbf{107}, (2023) no.7, 075032 
    [arXiv:2211.04595].
\bibitem{Imai:2022bke}
    H.~Imai, M.~Sakamoto, M.~Takeuchi and Y.~Tatsuta,
    ``Index and winding numbers on T2/ZN orbifolds with magnetic flux,''
    Nucl. Phys. B \textbf{990}, (2023) 116189 
    [arXiv:2211.15541].
\bibitem{Cremades} 
    D.~Cremades, L.~E.~Ibanez and F.~Marchesano,
    ``Computing Yukawa couplings from magnetized extra dimensions,''
    JHEP \textbf{05} (2004), 079
    [arXiv:hep-th/0404229].
\bibitem{Abe:2014vza}
    H.~Abe, T.~Kobayashi, K.~Sumita and Y.~Tatsuta,
    ``Gaussian Froggatt-Nielsen mechanism on magnetized orbifolds,''
    Phys. Rev. D \textbf{90}, (2014) no.10, 105006 
    [arXiv:1405.5012].
\bibitem{Fujimoto:2016zjs}
    Y.~Fujimoto, T.~Kobayashi, K.~Nishiwaki, M.~Sakamoto and Y.~Tatsuta,
    ``Comprehensive analysis of Yukawa hierarchies on $T^2/Z_N$ with magnetic fluxes,''
    Phys. Rev. D \textbf{94}, (2016) no.3, 035031 
    [arXiv:1605.00140].
\bibitem{Abe:2016eyh}
    H.~Abe, T.~Kobayashi, H.~Otsuka, Y.~Takano and T.~H.~Tatsuishi,
    ``Flavor structure in $SO(32)$ heterotic string theory,''
    Phys. Rev. D \textbf{94} (2016) no.12, 126020
    [arXiv:1605.00898].
\bibitem{Kobayashi:2016qag}
    T.~Kobayashi, K.~Nishiwaki and Y.~Tatsuta,
    ``CP-violating phase on magnetized toroidal orbifolds,''
    JHEP \textbf{04} (2017), 080 
    [arXiv:1609.08608].
\bibitem{Buchmuller:2017vut}
    W.~Buchmuller and K.~M.~Patel,
    ``Flavor physics without flavor symmetries,''
    Phys. Rev. D \textbf{97}, (2018) no.7, 075019 
    [arXiv:1712.06862].
\bibitem{Buchmuller:2017vho}
    W.~Buchmuller and J.~Schweizer,
    ``Flavor mixings in flux compactifications,''
    Phys. Rev. D \textbf{95}, (2017) no.7, 075024 
    [arXiv:1701.06935].
\bibitem{Neutrino} 
    M.~Ishida, K.~Nishiwaki and Y.~Tatsuta,
    ``Seesaw mechanism in magnetic compactifications,''
    JHEP \textbf{07} (2018), 125
    [arXiv:1802.06646].
\bibitem{Buchmuller} 
    W.~Buchmuller, M.~Dierigl, E.~Dudas and J.~Schweizer,
    ``Effective field theory for magnetic compactifications,''
    JHEP \textbf{04} (2017), 052
    [arXiv:1611.03798].
\bibitem{Ghilencea:2017jmh}
    D.~M.~Ghilencea and H.~M.~Lee,
    ``Wilson lines and UV sensitivity in magnetic compactifications,''
    JHEP \textbf{06} (2017), 039 
    [arXiv:1703.10418].
\bibitem{B1}
    W.~Buchmuller, M.~Dierigl and E.~Dudas,
    ``Flux compactifications and naturalness,''
    JHEP \textbf{08} (2018), 151
    [arXiv:1804.07497].
\bibitem{xi}
    T.~Hirose and N.~Maru,
    ``Cancellation of One-loop Corrections to Scalar Masses in Yang-Mills Theory with Flux Compactification,''
    JHEP \textbf{08} (2019), 054
    [arXiv:1904.06028].
\bibitem{Honda:2019ema}
    M.~Honda and T.~Shibasaki,
    ``Wilson-line Scalar as a Nambu-Goldstone Boson in Flux Compactifications and Higher-loop Corrections,''
    JHEP \textbf{03} (2020), 031 
    [arXiv:1912.04581].
\bibitem{Maru} 
    T.~Hirose and N.~Maru,
    ``Cancellation of One-loop Corrections to Scalar Masses in Flux Compactification with Higher Dimensional Operators,''
    J. Phys. G \textbf{48} (2021) no.5, 055005
    [arXiv:2012.03494].
\bibitem{Maru2}
    T.~Hirose and N.~Maru,
    ``Nonvanishing finite scalar mass in flux compactification,''
    JHEP \textbf{06} (2021), 159
    [arXiv:2104.01779].
\bibitem{Akamatsu}
    K.~Akamatsu, T.~Hirose and N.~Maru,
    ``Gauge symmetry breaking in flux compactification with a Wilson-line scalar condensate,''
    Phys. Rev. D \textbf{106} (2022) no.3, 035035
    [arXiv:2205.09320].
\bibitem{Maru:2023esr}
    N.~Maru and H.~Tanaka,
    ``Wilson-line Scalar Mass in Flux Compactification on an Orbifold $T^2/Z_2$,''
    arXiv:2303.01747.
\bibitem{Nielsen:1978rm}
    N.~K.~Nielsen and P.~Olesen,
    ``An Unstable Yang-Mills Field Mode,''
    Nucl. Phys. B \textbf{144}, (1978)  376-396.
\bibitem{Buchmuller:2019zhz}
    W.~Buchmuller, E.~Dudas and Y.~Tatsuta,
    ``Quantum corrections for D-brane models with broken supersymmetry,''
    JHEP \textbf{12} (2019), 022 
    [arXiv:1909.03007].
\bibitem{Buchmuller:2020nnl}
    W.~Buchmuller, E.~Dudas and Y.~Tatsuta,
    ``Tachyon condensation in magnetic compactifications,''
    JHEP \textbf{03} (2021), 070 
    [arXiv:2010.10891].
\bibitem{Haba:2002py}
    N.~Haba, M.~Harada, Y.~Hosotani and Y.~Kawamura,
    ``Dynamical rearrangement of gauge symmetry on the orbifold S1 / Z(2),''
    Nucl. Phys. B \textbf{657} (2003), 169-213
    [erratum: Nucl. Phys. B \textbf{669} (2003), 381-382]
    [arXiv:hep-ph/0212035].
\bibitem{Haba:2003ux}
    N.~Haba, Y.~Hosotani and Y.~Kawamura,
    ``Classification and dynamics of equivalence classes in SU(N) gauge theory on the orbifold S**1 / Z(2),''
    Prog. Theor. Phys. \textbf{111} (2004), 265-289
    [arXiv:hep-ph/0309088].
\bibitem{Kawamura:2022ecd}
    Y.~Kawamura, E.~Kodaira, K.~Kojima and T.~Yamashita,
    ``On representation matrices of boundary conditions in SU(n) gauge theories compactified on two-dimensional orbifolds,''
    JHEP \textbf{04} (2023), 113
    [arXiv:2211.00877].
\bibitem{tHooft:1979rtg}
    G.~'t Hooft,
    ``A Property of Electric and Magnetic Flux in Nonabelian Gauge Theories,''
    Nucl. Phys. B \textbf{153} (1979), 141-160.
\bibitem{Hall:2001tn}
    L.~J.~Hall, H.~Murayama and Y.~Nomura,
    ``Wilson lines and symmetry breaking on orbifolds,''
    Nucl. Phys. B \textbf{645} (2002), 85-104
    [arXiv:hep-th/0107245].
\bibitem{T1}
    D.~Tong,
    ``Lectures on the Quantum Hall Effect,''
    arXiv:1606.06687.
\bibitem{Scrucca:2003ut}
    C.~A.~Scrucca, M.~Serone, L.~Silvestrini and A.~Wulzer,
    ``Gauge Higgs unification in orbifold models,''
    JHEP \textbf{02} (2004), 049
    [arXiv:hep-th/0312267].
\bibitem{nonflat5D}
  K.~Kojima, K.~Takenaga and T.~Yamashita,
  ``Multi-Higgs Mass Spectrum in Gauge-Higgs Unification,''
  Phys. Rev. D \textbf{77}, (2008), 075004
  [arXiv:0801.2803].
\bibitem{Alfaro:2006is}
    J.~Alfaro, A.~Broncano, M.~B.~Gavela, S.~Rigolin and M.~Salvatori,
    ``Phenomenology of symmetry breaking from extra dimensions,''
    JHEP \textbf{01} (2007), 005 
    [arXiv:hep-ph/0606070].
\bibitem{E6twist}
    M.~Bando and T.~Kugo,
    ``Neutrino masses in E(6) unification,''
    Prog. Theor. Phys. \textbf{101} (1999), 1313-1333
    [arXiv:hep-ph/9902204].
\bibitem{Bando:2000gs}
    M.~Bando, T.~Kugo and K.~Yoshioka,
    ``Mass matrices in E(6) unification,''
    Prog. Theor. Phys. \textbf{104} (2000), 211-236
    [arXiv:hep-ph/0003220].
\bibitem{Inoue:2007ed}
    K.~Inoue, K.~Kojima and K.~Yoshioka,
    ``Probing flavor structure in unified theory with scalar spectroscopy,''
    JHEP \textbf{07} (2007), 027 
    [arXiv:hep-ph/0703253].



\end{thebibliography}
\end{document}